\documentclass[sigconf]{acmart}

\usepackage{times}
\usepackage{epsfig}
\usepackage{graphicx}
\usepackage{amsmath}
\usepackage{subfigure}
\usepackage{amsthm}
\usepackage[section]{placeins}
\usepackage{svg}
\usepackage{float}

\renewcommand\footnotetextcopyrightpermission[1]{}

\newtheorem{myDef}{Definition}
\newtheorem{myTheo}{Theorem}

\AtBeginDocument{%
  }

\acmConference[]{}{}

\begin{document}

\title{FedSOV: Federated Model Secure Ownership Verification with Unforgeable Signature}

\author{Wenyuan Yang}
\email{yangwy56@mail.sysu.edu.cn}
\affiliation{%
  \institution{Sun Yat-sen University}
  \city{Shenzhen}
  \country{China}
}

\author{Gongxi Zhu}
\authornote{This work is done by Gongxi Zhu and Yuguo Yin during their internship at Sun Yat-sen University, under the guidance of Wenyuan Yang.}
\email{gx.zhu@foxmail.com}
\affiliation{%
  \institution{University of Electronic Science and Technology of China}
  \city{Chengdu}
  \country{China}
}

\author{Yuguo Yin}
\authornotemark[1]
\email{yuguoyin2002@gmail.com}
\affiliation{%
  \institution{University of Electronic Science and Technology of China}
  \city{Chengdu}
  \country{China}
}

\author{Hanlin Gu}
\email{ghltsl123@gmail.com}
\affiliation{%
  \institution{WeBank}
  \city{Shenzhen}
  \country{China}
}

\author{Lixin Fan}
\authornote{Lixin Fan is the corresponding author.}
\email{lixinfan@webank.com}
\affiliation{%
  \institution{WeBank}
  \city{Shenzhen}
  \country{China}
}

\author{Qiang Yang}
\email{qyang@cse.ust.hk}
\affiliation{%
  \institution{The Hong Kong University of Science and Technology}
  \city{Hong Kong}
  \country{China}
}

\author{Xiaochun Cao}
\email{caoxiaochun@mail.sysu.edu.cn}
\affiliation{%
  \institution{Sun Yat-sen University}
  \city{Shenzhen}
  \country{China}
}

\renewcommand{\shortauthors}{Yang et al.}

\pagestyle{plain}

\begin{abstract}
Federated learning allows multiple parties to collaborate in learning a global model without revealing private data. The high cost of training and the significant value of the global model necessitates the need for ownership verification of federated learning. However, the existing ownership verification schemes in federated learning suffer from several limitations, such as inadequate support for a large number of clients and vulnerability to ambiguity attacks. To address these limitations, we propose a cryptographic signature-based federated learning model ownership verification scheme named FedSOV. FedSOV allows numerous clients to embed their ownership credentials and verify ownership using unforgeable digital signatures. The scheme provides theoretical resistance to ambiguity attacks with the unforgeability of the signature. Experimental results on computer vision and natural language processing tasks demonstrate that FedSOV is an effective federated model ownership verification scheme enhanced with provable cryptographic security.
\end{abstract}





\maketitle

\section{Introduction}
\label{Sect:Introduction}
Training a deep learning model requires a lot of resources, including 1) high-performance professional hardware and complex model design as the basis for model training \cite{wang2021voxpopuli}; 2) a large amount of training data and long training time to improve the performance of the model \cite{deng2009imagenet,  chelba2013one, zhu2015aligning}.
For the BERT model with 1.5 billion parameters, the training cost is between \$80,000 and \$1.6 million \cite{sharir2020cost}, and GPT-3 has a compute cost of \$46,000 per training run \cite{dale2021gpt}. There is a compelling need to protect these increasingly larger and more costly models from being plagiarized.
This is especially true for Federated Learning (FL) which is a privacy-preserving training paradigm that collaboratively trains federated models without sharing client data. It is necessary to prevent adversaries from plagiarizing, misusing, and redistributing valuable FL models without the legal permission of the legitimate model owners \cite{yang2023federated}. 

Existing federated model ownership verification (FMOV) schemes such as FedIPR \cite{li2022fedipr} and WAFFLE \cite{tekgul2021waffle} have been proposed. They embed watermarks into models as ownership credentials and verify the model ownership  by extracting the watermark.
However, these schemes suffer from two significant limitations, which hinder their application in certain scenarios.
First, the effectiveness of embedded watermarks decreases significantly with an increasing number of clients requiring to embed their credentials. As evidenced by the decreased watermark detection accuracy in Fig. \ref{Fig:ClientNum}, these FMOV schemes are unsuitable for federated systems with a large number of clients.
Second, the watermark is embedded directly into the model as a credential to declare ownership. Adversaries may launch the \emph{ambiguty attack} \cite{fan2019rethinking} by extracting their own forged credentials from the model and rendering legitimate ownership verification doubtful or useless.

To deal with the above concerns, we propose a new FMOV scheme called FedSOV.
First, FedSOV is suitable for realistic complex FL systems with a large number of clients by compressing the public keys of clients into a short watermark with the hash function. This scheme ensures that the effectiveness of embedded watermarks is not affected by the number of clients (see Sect. \ref{Sect:Watermark Generation}).
Second, FedSOV resists the ambiguity attack by incorporating unforgeable digital signature verification (see Sect. \ref{Sect:AmbiguityAttack}), in which a cryptographic hard problem  provably rules out the possibility of attackers forging a valid signature to claim ownership.

Moreover, the application of digital signature in FMOV poses a technical challenge due to \emph{watermark errors}. In particular, the errors introduced by federated model training and watermark removal attacks can impact the correctness of ownership verification. To overcome this challenge, we propose the use of near-collision resistant hash and provide theoretical analysis in Section 6.2 regarding the security boundary for FedSOV. Our analysis reveals that even if the detection rate of a 2048-bit watermark drops to 82.1\%, ownership can still be securely verified with overwhelming probability in Sect. \ref{Sect:CombinationAttack}. Experimental results presented in Sect \ref{Exp:robustness} verify the robustness of FedSOV against removal attacks, with watermark detection accuracy above the secure boundary.

Our contributions are as follows:
\begin{itemize}
    \item We propose a model watermarking scheme that allows a large number of clients to embed their own ownership credentials, enabling larger federated learning systems than previous schemes (see Sect. \ref{Sect:Method}).
    \item We use the digital signature for the first time in FMOV, which effectively resists the ambiguity attack with provable cryptographic security (see Sect. \ref{Sect:AmbiguityAttack}).
    \item Moreover, we solve the challenge of \emph{watermark error} by near-collision resistance hash and provide theoretical security boundary (see Sect. \ref{Sect:CombinationAttack}) which is empirically validated with extensive experimental validation (see Sect. \ref{Exp:robustness}).  
\end{itemize}

\section{Related Work}
\label{Sect:RelatedWork}
\subsection{Federated Learning}
Federated learning is originally proposed for multiple parties to collaboratively train machine learning models without sharing private data \cite{mcmahan2016federated,mcmahan2017communication, konevcny2016federated}. Furthermore, Yang et al. \cite{yang2019federated} expand FL according to different types of cross-institutional data utilization problems and divided them into three categories:\textit{horizontal federated learning}, \textit{vertical federated learning} and \textit{federated transfer learning}. Recently, Yang et al. \cite{yang2023federated}  propose \textit{secure federated learning}, in which the importance of model intellectual property rights is emphasized.

\subsection{DNN Model Watermarking}
The intellectual property protection of deep learning models is mainly realized through model watermarking. Existing model watermarking schemes can be divided into two categories: feature-based watermarking \cite{uchida2017embedding,fan2019rethinking,fan2021deepip,chen2018deepmarks,zhang2020passport} and backdoor-based watermarking \cite{adi2018turning,zhang2018protecting,lukas2019deep}. Since backdoor-based watermarking is limited to classification tasks and is in itself a threat of model poisoning, our scheme adopts feature-based watermarking.

For property protection in federated learning models, Tekgul et al. \cite{tekgul2021waffle} proposes WAFFLE, which adds a retraining step to embed the watermark after the local model is aggregated on the server side, but does not allow the client to embed and verify the watermark. Liu et al. \cite{liu2021secure} propose a backdoor-based scheme where the client is responsible for embedding a watermark in a homomorphically encrypted FL framework. Another representative solution is FedIPR \cite{li2022fedipr} which allows each client to embed its own watermark in the model. However, the above-mentioned property protection systems all use watermarks as ownership certificates, and the verification process leaks the certificates, and the security cannot be proved.

\subsection{Vulnerability of DNN Watermarking }

Watermark is a key certificate for ownership verification, and attackers can interfere with and disrupt normal ownership verification through a series of attacks against watermarks. Since this solution is based on feature-based watermarking,  here are the security vulnerabilities faced by  feature-based watermarking. These attacks mainly fall into two categories:

\textbf{Ambiguity Attack: }The ambiguity attack originates from image watermarking,  which constructs a forged watermark with an inverted process to confuse ownership during verification \cite{craver1998resolving,li2006zero,sencar2007combatting}. In the field of model watermarking, Fan et al. \cite{fan2019rethinking} first proposed the ambiguity attack. An ambiguity attacker can modify the watermarked model with minor computational and without original training datasets, so that an additional reverse-engineered watermark can be detected from the model (defined in Def. \ref{Def:Ambiguity Attack}). Fan et al. also proved by experiments that none of the existing watermarking methods in \cite{fan2019rethinking} are able to deal with ambiguity attacks. 
Wang et al. \cite{wang2022rethinking} point out from the attack condition that many watermark schemes are difficult to meet the requirements of the ambiguity attack, but FedSOV can resist this attack theoretically.

In order to resist the ambiguity attack, Fan et al. \cite{fan2019rethinking} designed the passport layer, through which the function of the model is controlled by the passport, but Chen et al. \cite{chen2023effective} designed an advanced ambiguity attack against the passport-based method, which can forge multiple valid passports with a small training dataset. FedSOV does not adopt the passport-based method, but overcomes this attack theoretically based on digital signatures.  In the field of image watermarking, there are also attempts to use cryptographic signatures to resist ambiguity attacks \cite{adelsbach2003watermarking}, but they cannot solve the problem that the errors of detection lead to the failure of cryptographic constructions. FedSOV uses near-collision to solve a similar error problem in model watermarking.

\textbf{Removal Attack: }The attacker tries to remove the watermark embedded in the model while maintaining the model performance as much as possible. The main techniques of removal attacks include model fine-tuning and pruning, and many attack methods based on these two techniques have been derived. Wang et al. \cite{wang2022rethinking} proposed a new stealthy and powerful watermark removal attack against backdoor-based watermarking, which disrupts the embedded watermarks via injecting relighting perturbations in total preprocessing. FedSOV is based on feature-based watermarking, so this attack is not considered.
Following the previous DNN model watermarking methods \cite{fan2021deepip,zhang2021deep,li2022fedipr}, we mainly study the robustness of the proposed watermark against removal attacks under pruning and fine-tuning.

\subsection{Cryptographic Digital Signature}
Digital signature has the following functions: data integrity and data origin identification. Data integrity means that signed data cannot be changed by unauthorized means. Data origin identification means that the data sender is the same as claimed. Digital signature schemes have been used as primitives in cryptographic protocols \cite{signature:blake1997entity, signature:turner2014transport, signature:doraswamy2003ipsec} that provide entity authentication and authenticated key transport. As the research progresses, there are various digital signature schemes, including integer  factorization problem-based ones \cite{signature:rivest1978method}, discrete logarithm problem-based ones \cite{signature:elgamal1985public,signature:johnson2001elliptic, signature:abram2022low}, difficult problems on lattice-based ones \cite{signature:espitau2022shorter,signature:beullens2023group}, etc. FedSOV implements FMOV using the data origin identification of signature.
\section{Preliminaries}
\label{Sect:Preliminaries}

\subsection{Notations}
The bold lower-case and upper-case letters like $\mathbf{a, A}$ denote vectors and matrices. The $\{0,1\}^n$ is the set of binary strings of length $n$, and $\{0,1\}^*$ is the set of binary strings with arbitrary length. The Hamming weight of $\mathbf{a}\in\{0,1\}^{m}$ is denoted as $||\mathbf{a}||_1=\sum^{m}_{i=1}\mathbf{a}[i]$. The Hamming distance between $\mathbf{a,b}\in\{0,1\}^m$ is computed as $\sum^{m-1}_{i=0}(\mathbf{a}_i\oplus\mathbf{b}_i)$. The concatenation symbol is $||$. The main notations used in this paper are shown in Tab. \ref{Tab:MainNotations}.

\begin{table}[]
    \centering
    \caption{Main Notations Used in This Paper}
    \begin{tabular}{|c|c|}
    \hline
       \textbf{Notations}  & \textbf{Descriptions} \\
       \hline
        $K$ & The number of clients in FL\\
        $\mathbb M$ &The DNN model\\
        $\rm{H}(\cdot)$ & The hash function\\
        $n$ & The output length of the hash function\\
        $\mathbf{E}$ & The watermark embedding matrix\\
        $pk$ & The public key of signature\\
        $sk$ & The private key of signature\\
        $\sigma$ & The signature for verification\\
        $pk_{con}$ & The concatenated public key\\
        $\mathbf{h}$ & The hash watermark of FedSOV\\
        $\mathbf{B}$ & The common watermark of previous work\\
        $\mathbf{D}$ & The training data sets\\
        $\mathbf {W}_{t}$ & The target model parameters embedding watermarks\\
        $\mathbf {W}_{\gamma}$ & Weights of normalization layer\\
        $r(n)$ &  Security Boundary of Watermark Detection Rate\\
        \hline
    \end{tabular}
    \label{Tab:MainNotations}
\end{table}

\subsection{Cryptographic Digital Signature}
\label{Sect:DigitalSignature}
A digital signature scheme \cite{signature:boneh2004short} includes four algorithms, $Setup$, $KeyGen$, $Sign$ and $Verify$. 
\begin{enumerate}
    \item $Setup(1^{\lambda})\rightarrow SP$: It uses the input of security parameter $\lambda$ to initialize the system parameters $SP$ of the digital signature system.
    \item $KeyGen(SP)\rightarrow(pk,sk):$ It inputs $SP$, and outputs public key $pk$ and private key $sk$. The $pk$ is public to everyone, the $sk$ can only be held by the owner of the $pk$.
    \item $Sign(m,sk)\rightarrow \sigma$: It inputs message $m\in\{0,1\}^{*}$ and private key $sk$, and outputs signature $\sigma$.
    \item $Verify(m,\sigma,pk)\rightarrow 1/0$: It inputs message $m$, signature $\sigma$ and public key $pk$. If the signature verification passes, the algorithm outputs 1, otherwise, it outputs 0.
\end{enumerate}

The identification of a signature $\sigma$ can be understood as "the user with private key $sk$ of public key $pk$ has sent message $m$". It is feasible to construct an ownership verification scheme using the identification function of the digital signature.

\subsection{Hash Near-collision}
Given two binary strings $\mathbf{a},\mathbf{b}\in\{0,1\}^{n}$. A function $\rm{Dif}(\mathbf{a},\mathbf{b})=\sum^{n}_{i=1}(\mathbf{a}[i]\oplus\mathbf{b}[i])$ denotes the Hamming distance between $\mathbf{a},\mathbf{b}$.

Let $\mathcal{H}=\{\rm{H}:\{0,1\}^{*}\rightarrow\{0,1\}^{n}\}$ be a family of hash functions. For $n'\in\{1,...,n\}$ and $\mathbf{x},\mathbf{x}'\in\{0,1\}^{*}$, we say that $n'$-near-collision \cite{hash:biham2004nearcollisionSHA0} happens between $\mathbf{x}$ and $\mathbf{x}'$ if the Hamming distance between $Dif(\rm{H}(\mathbf{x}),\rm{H}(\mathbf{x}'))\leq n'$.
\section{PROBLEM DEFINITION}
\label{Sect:PROBLEM DEFINITION}

\subsection{System Model}
The system model focuses on the role of each entity in FMOV. There are three entities in FedSOV including client, server, and verifier.  \textbf{Clients} build their own local models using their own datasets, send their own public keys to the server to get the watermark, and embed the watermark into the local models. The \textbf{server} generates the watermark from clients' public keys and aggregates the clients' local models into a global model. The \textbf{verifier} verifies ownership of models suspected of being stolen. He is an honest but curious entity outside the FL system responsible for deriving ownership verification results. The identity of the verifier varies depending on the scenario such as courts, inspectorates, notary offices, or users who want to confirm whether the service provider is the legitimate owner of the model, etc.

\subsection{Threat Model}


The threat model focuses on the trust assumptions about the entities in FMOV and the capabilities of the attacker. In FMOV, we have the following trust assumptions about these three types of entities:
\begin{itemize}
    \item Clients are trusted, they perform all the steps that require their action correctly to preserve their ownership interests.
    \item Server is honest but curious. The server properly executes watermark generation and federal aggregation. However, the server does not have model ownership in our settings because he did not contribute private data and other enough precious training resources.
    \item The verifier is honest but curious. The verifier performs the ownership verification process correctly, but tries to get the private credential of clients.
\end{itemize}

The attacker can be the entity inside or outside the FL system including the server, verifier, and other entities. He aims to use his capabilities to pass ownership verification on a model that does not belong to him. The capabilities an attacker can obtain from FMOV are listed as follows:
\begin{itemize}
    \item \textbf{Wiretapping}: The attacker can obtain the clients' public keys and signatures in previously public ownership verification.
    \item \textbf{Ambiguity Attack}: The attacker tries to use his eavesdropped public keys and signatures to forge a new valid signature that can pass ownership verification (defined in Def. \ref{Def:Ambiguity Attack against FedSOV}).
    \item \textbf{Removal Attack}: After the attacker steals some models, he can modify the watermarks of the stolen model by the removal attack. However, breaking the watermark reduces the model's accuracy inevitably \cite{fan2021deepip,li2022fedipr}. Therefore, to obtain a model with acceptable accuracy, he cannot break the watermark on a large scale.
\end{itemize}
\section{Proposed Scheme}
\label{Sect:Method}
\begin{figure*}
    \centering
    \includegraphics[scale=0.5]{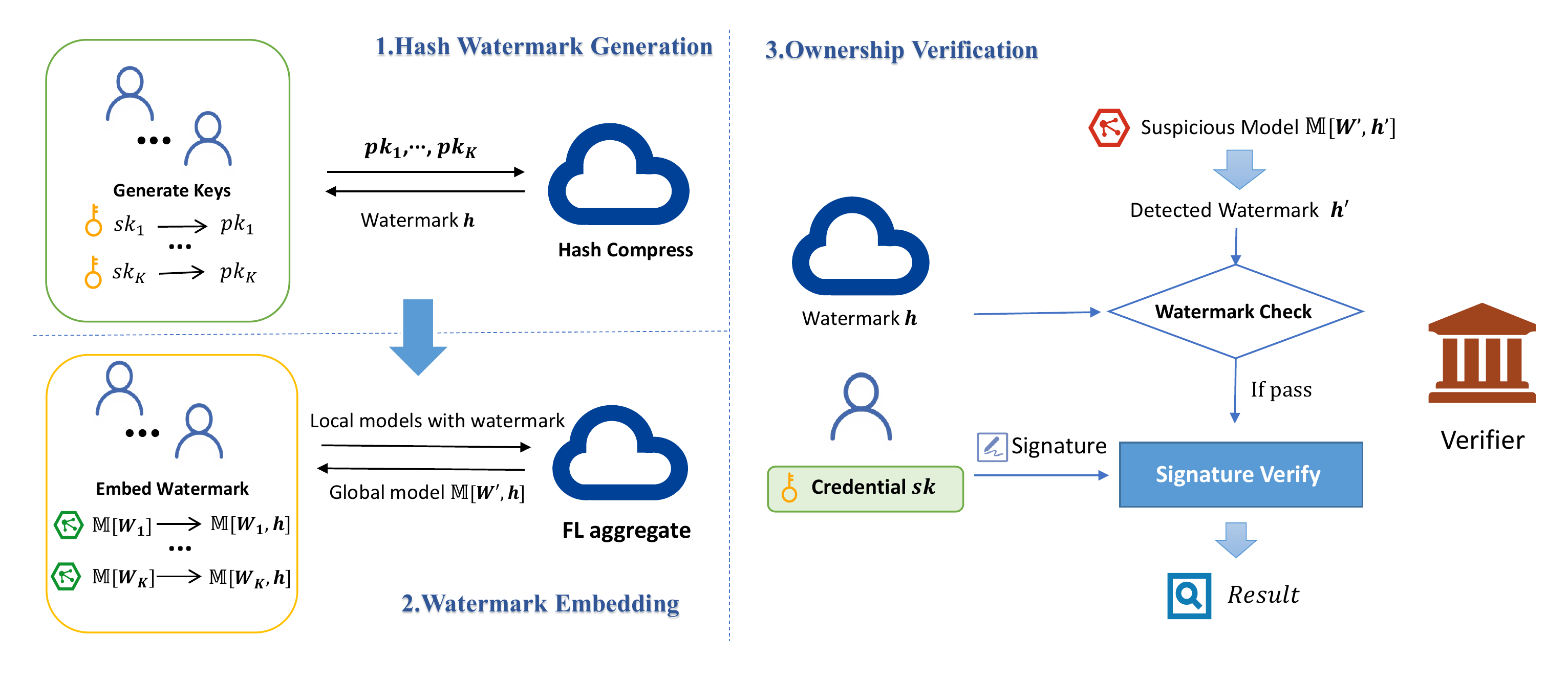}
    \caption{An illustration of FedSOV. First, each client uses its own private key $sk$ to generate a public key $pk$, and then sends it to the server to calculate the global hash watermark $\mathbf h$. Second, each client performs the watermark embedding process, embeds $\mathbf h$ into the model $\mathbb {M}[\mathbf {W}_i]$, and uploads the local model to the federated aggregation server to obtain the global model containing the watermark $\mathbb {M}[\mathbf {W'},\mathbf h]$. Third, when a model is suspected of being plagiarized, each client can perform ownership verification with the verifier. The verifier compares the extracted watermark to check the watermark, and then if it passes, he performs signature verification to check whether the client has the ownership credential $sk$ or not.}
    \label{Fig:SystemModel}
\end{figure*}
This section outlines the four steps involved in the FedSOV process: \textit{System Setup}, \textit{Hash Watermark Generation}, \textit{Watermark Embedding}, and \textit{Ownership Verification}. First, the system's public parameters are initialized in Sect. \ref{Sect:SystemSetup}.
Then during the hash watermark generation process in Sect. \ref{Sect:WatermarkGen}, each client generates its own private key as the ownership credential. The public key is concatenated by the server through the hash function to obtain a global common watermark and distribute it to each client. This generation scheme compresses public keys  into an equal-length watermark such that even if a huge federated system has many clients, FedSOV can still support each client to use its own credentials to verify ownership without damage to the watermark embedding effect, which is almost impossible for the previous schemes such as FedIPR. After that, we adopt the watermark embedding method in \cite{li2022fedipr} to embed the global watermark in Sect. \ref{Sect:Watermark Embedding}. Finally, for the ownership verification scheme in Sect. \ref{Sect:Ownership Verification}, each client can use its own private key to publicly prove the ownership of the model to the verifier through the signature verification algorithm. The unforgeability of signatures ensures that even if an attacker wiretaps the verification process and obtains some public keys and signatures, he cannot successfully launch the ambiguity attack to forge a new valid signature.
An illustration of FedSOV is shown in Fig.\ref{Fig:SystemModel}.
\subsection{System Setup}
\label{Sect:SystemSetup}
This step sets up the public parameters $PP$ used in this system. The $PP$ is the standard for this system that can be published by a standards institute, which is constructed as 
\begin{equation}\nonumber
    \begin{aligned}
        (\mathcal{P},\mathbf{E},SP,\rm{H}(\cdot), r(n)).
    \end{aligned}
\end{equation}
\begin{itemize}
    \item $\mathcal{P}$ is a positional parameter that determines which layers of the model the watermark is embedded in.
    \item $\mathbf{E}$ is the embed matrix and $\mathbf E^{\omega \times n} {\leftarrow}\mathcal{N}(0,1)$ in which $\omega$ is the number of model parameters determined by $\mathcal{P}$, $n$ is the length of the watermark and $\mathcal{N}(0,1)$ is the standard normal distribution.
    \item $SP$ is the system parameter of the digital signature scheme generated by algorithm $Setup(1^{\lambda})$.
    \item $\rm{H}(\cdot)$, the near-collision resistance hash that maps the public key to a binary string with length $n$.
    \item $r(n)$ (see Def. \ref{Def:Security Boundary}) is the security boundary of the watermark detection rate (see Def. \ref{def:watermark-detect}). It is a reasonable premise to address secure technical challenges, i.e., \emph{watermark error}.
\end{itemize}

\subsection{Hash Watermark Generation}
\label{Sect:Watermark Generation}

\subsubsection{Insight}

In previous FedMOV schemes such as FedIPR\cite{li2022fedipr}, the watermark embedded into the model is the only model ownership credential, which leads to an increase in the length of the watermark when the number of clients increases. As the number of model parameters responsible for containing the embedded information is limited, the amount of credential information that can be embedded is limited. \textit{When the number of clients exceeds the threshold, the watermark is too long to be effectively embedded}. This limit is disclosed by the analysis of Theorem 1 in \cite{li2022fedipr}.

Inspired by the hash function, which compresses a message of any length into a fixed-length message digest, we design a watermark generation scheme that compresses the credential information of all clients into a global watermark through the hash function. No matter how many clients there are, the length of the watermark embedded into the model is always fixed, which enables our scheme to be adopted to federated systems with a huge number of clients. The comparison analysis between FedSOV and FedIPR on this feature can be seen in Experiment \ref{Exp:EmbeddingEffect}.

\subsubsection{Method to Generate Hash Watermark}
\label{Sect:WatermarkGen}

For each client $i\in\{1,...,K\}$ in the federated system, it runs $KeyGen(SP)$ algorithm in Sect. \ref{Sect:DigitalSignature} to get public key $pk_i$ and private key $sk_i$. Then each client sends its public key $pk_i$ to the server.

After the server obtains everyone's public key, it uses the hash function to calculate the hash watermark $\mathbf h$ as follows: 
\begin{equation}
    \label{Equ.Hash}
    \mathbf h={\rm H} (\ pk_{con}),
\end{equation}
in which $pk_{con}=pk_1||pk_2||\cdots||pk_K$, $\rm {H}$$(pk_{con})$ is the hash watermark.
The hash function in FedSOV can be implemented by SHAKE algorithm \cite{dworkin2015sha} of SHA3. 
    
Finally, $\mathbf h$ is distributed by the server to all clients as the watermark to be embedded into the model.

It should be pointed out that no matter how long the input is, $\rm{H}(\cdot)$ will always map it to an output of fixed length n. Therefore, no matter how many clients there are, this scheme can compress the ownership credential information of the client into a fixed-length watermark, thereby avoiding the decrease of the watermark embedding effect caused by the increase of client number.

\subsection{Watermark Embedding}
\label{Sect:Watermark Embedding}

$ \mathbf W_t$ is the target model parameters to be embedded in the watermarks, which come from one or more parts of a model and its position can be flexibly selected by the positional parameter $\mathcal{P}$ . The embedding process is done by adding the following loss regularization term into the main loss :
\begin{equation}
\begin{split}
{L}_{h}(\mathbf{W}_t \mathbf{E}, \mathbf h)&=\alpha \rm{HL}(\mathbf h', \mathbf h)\\\
&=\alpha \sum\limits^n\limits_{i=1} \rm {max}(\mu - b_i t_i),
\end{split}
\end{equation}
whereas $\alpha$ denotes the relative weight factor to control the embedding loss (we set $\alpha=0.5$), $\rm {HL}()$ is a hinge-like loss to implement the regularization term and  $\mu$ is the parameter of hinge loss, $$\mathbf h' = \rm{sgn}(\mathbf{W}_t \mathbf{E})=(b_1,\cdots,b_n)\in \{0,1\}^n$$ is the extracted watermarks, $\mathbf h=(t_1,\cdots,t_n) \in \{0,1\}^n $ is the target watermark, and  $\rm{sgn()}$ is a sign function.

According to the structure of a convolution layer, we can choose the convolution kernel weights $\mathbf W_C$ or the normalization layer weights $\mathbf W_{\gamma}$ as the target model parameters $\mathbf W_t$. Fan et al.\cite{fan2021deepip} find that embedding  watermarks into the  normalization layer weights is more conducive to the robustness of  watermarks. Therefore,  FedSOV embeds the  watermarks into the normalization layer scale parameters of the convolution block, i.e., $ \mathbf W_t=\mathbf W_{\gamma}=(\gamma^{(1)},\cdots,\gamma^{(C)})$. The normalization layer is implemented as follows:
\begin{equation}
\begin{split}
\mathbf y^{(i)}
&=\mathbf \gamma ^{(i)} \ast \mathbf x^{(i)} +\mathbf\beta^{(i)},
\end{split}
\end{equation}
where $\mathbf \gamma^{(i)}$is the scale weight in channel $i$ ,  $\mathbf x^{(i)} $ is the input of the normalization layer and $\mathbf\beta^{(i)}$ is the offset parameter of  normalization.

\subsection{Ownership Verification}
\label{Sect:Ownership Verification}
When a model $\mathbb{M}(\mathbf{W}', \mathbf{h}')$ is suspected of being stolen from the FL system, its ownership can be verified through the ownership verification process.
As opposed to directly using the watermark as a credential to verify ownership, FedSOV uses the private key as the credential and proves ownership with a signature. Due to the unforgeability of the digital signature, it is difficult to forge a valid signature even if an attacker obtains a series of public keys and signatures, which successfully resists ambiguity attacks. The specific analysis of this feature is shown in Sect. \ref{Sect:AmbiguityAttack}.


Due to the technical challenge that watermark error induced by model training and removal attacks may affect the correctness of verification, FedSOV performs watermark checks as shown in Sect. \ref{Sect:CombinationAttack} to overcome this security problem.


Then if the watermark check passes, the client $c_i$ uses its private key $sk_i$ to  execute \emph{signature verification} with the verifier.
The signature verification is as follows:

\begin{itemize}
    \item First, the verifier randomly chooses a message $m$, and sends it to $c_i$. 
    \item Then the client $c_i$ executes the signature algorithm $Sign(m,sk_i)$ to get the signature $\sigma$, and sends $\sigma$ to the verifier.
    \item Finally, the verifier executes signature verification algorithm $Verify(m,\sigma,pk_i)$ and return the verification result.
\end{itemize}

The client $c_i$ can pass ownership verification and claim the ownership of the suspicious model only if $Verify(m,\sigma,pk_i)$ outputs 1. FedSOV adopts \emph{short signature algorithm} \cite{signature:boneh2004short} shown in Appendix \ref{Appe:ShortSignature} to implement signature verification.
\section{Security Analysis}
\label{Sect:SecurityAnalysis}
\subsection{Security Against Ambiguity Attack}
\label{Sect:AmbiguityAttack}

This section presents the ability of the proposed FedSOV to withstand ambiguity attacks (Theorem \ref{Theo:SDH}), while highlighting the inability of FedIPR \cite{li2022fedipr} to resist such attacks(Theorem \ref{Theo:FedIPR-AA}). The proof and example are deferred in Appendix \ref{Apdix:Proof1}. 

\subsubsection{Vulnerability of FedIPR against Ambiguity Attack}

The previous FMOV schemes like FedIPR \cite{li2022fedipr} use watermark as the credential to claim ownership. An attacker can extract a forged and valid watermark that can declare model ownership without changing the model parameters.

Following \cite{fan2019rethinking}, when the watermark is used as the ownership verification credential, the ambiguity attack can be defined as follows:

\begin{myDef}[Ambiguity Attack]
\label{Def:Ambiguity Attack}
For a model $\mathbb M$ with weights $\mathbf W$  and common watermark $\mathbf B$ i.e.,  $\mathbb M [\mathbf W, \mathbf B]$, an ambiguity attack can be successfully constituted if
\begin{itemize}
    \item [(1)] a forged watermark $\mathbf B'$ can be reverse-engineered for a given DNN model $\mathbb M$;

    \item [(2)] the forged $\mathbf B'$ can be successfully verified ownership with respect to the given DNN weights $\mathbf W$;

    \item [(3)] the discrepancy of model inference performance is less than a predefined threshold i.e., $|\mathcal F(\mathbb M[\mathbf W,\mathbf B']), \mathbf {D}_t)-\mathcal {F}_t| \leq \epsilon _f$, in which $\mathcal F(\mathbb M[\mathbf W,\mathbf B']),\mathbf {D}_t)$ is the model inference performance tested on test data $\mathbf {D}_t$, $\mathcal {F}_t$ is the target inference performance. 
\end{itemize}
\end{myDef}

\begin{myTheo}
\label{Theo:FedIPR-AA}
For a targeted forged watermark $\mathbf {B'}=(b'_1,\cdots,b'_n)$ and targeted model parameters $\mathbf {W}_t=(w_1,w_2,\cdots,w_{\omega})$, it is easy to  find a forged embedding matrix $\mathbf E'=\{e_{ij}\}_{\omega \times n}$ for \textbf{FedIPR} that satisfies $\rm sgn (\mathbf {W}_t \mathbf {E'} )=\mathbf B'$, i.e.,

\begin{equation}
\left\{\begin{array}{c}    (w_1e_{11}+w_2e_{21}+\cdots+w_{\omega}e_{\omega 1})b'_1>0 \\    (w_1e_{12}+w_2e_{22}+\cdots+w_{\omega}e_{\omega 2})b'_2>0 \\    \cdots \\     (w_1e_{1n}+w_2e_{2n}+\cdots+w_{\omega}e_{\omega n})b'_n>0\end{array}\right. .
\end{equation}
The three conditions in Def.\ref{Def:Ambiguity Attack} are all satisfied. An ambiguity attack can be successfully constituted for FedIPR models.

\end{myTheo}

 Theo. \ref{Theo:FedIPR-AA} means that an ambiguity attacker can easily extract his forged watermark from the FedIPR model without changing the model parameters and claim his ownership.


\subsubsection{Security of FedSOV against Ambiguity Attack}

In FedSOV, ownership verification is performed by means of signature verification, and the watermark is a hash-compressed digest of the public key of the signature algorithm. Thus the ambiguity attacker's goal changes from forging watermarks to forging valid signatures. Therefore, the ambiguity attack against FedSOV can be defined as follows:
\begin{myDef}[Ambiguity Attack against FedSOV]
\label{Def:Ambiguity Attack against FedSOV}
For a model $\mathbb M$ with weights $\mathbf W$  and public hash watermark $\mathbf h$ i.e.,  $\mathbb M [\mathbf W, \mathbf h]$,  the ownership can be verified through signature $\sigma$. An ambiguity attack against FedSOV can be successfully constituted if
\begin{itemize}
    \item [(1)] a new signature $\sigma '$ can be constructed from a given model $\mathbb N$ and public information including $\mathbf h$;
    \item[(2)] the signature $\sigma '$ is valid, through which the attacker can pass the ownership verification;
    \item[(3)] the discrepancy of model inference performance is less than a predefined threshold,i.e., $|\mathcal M(\mathbb M[\mathbf W,\mathbf h]) ,\mathbf {D}_t)-\mathcal {M}_t| \leq \epsilon _f$.
\end{itemize}
\end{myDef}

\begin{myTheo}
\label{Theo:SDH}
For any probabilistic polynomial attacker, if an attacker can successfully forge a new valid digital signature of FedSOV with a non-negligible advantage, then there exists a probabilistic polynomial-time algorithm that can solve the q-SDH problem with a non-negligible advantage.
\end{myTheo}

\begin{figure}
    \flushleft 
    \includegraphics[scale=0.4]{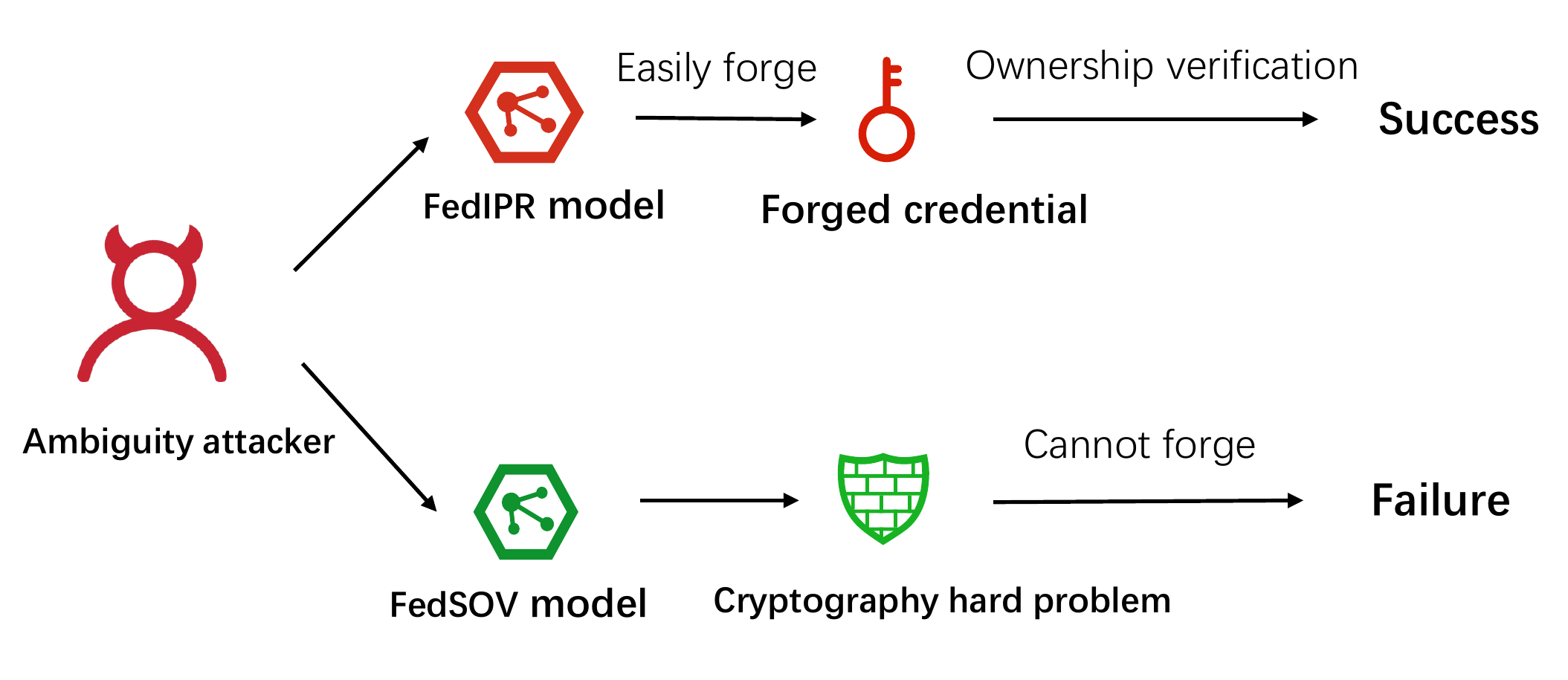}
    \caption{An ambiguity attacker can easily forge FedIPR's ownership credential, but cannot forge FedSOV's credential with cryptography hard problem guarantees.}
    \label{Fig:CompareAA}
\end{figure}

Therefore, Theo. \ref{Theo:SDH} demonstrates short signature used in FedSOV is unforgeable under the $q$-SDH problem \cite{signature:boneh2004short}. Specifically, even if an attacker intercepts a series of public keys and signatures, he can only successfully forge a new valid signature with negligible probability. As a result, condition (2) in Def. \ref{Def:Ambiguity Attack against FedSOV} is not satisfied, which implies FedSOV is theoretically secure against ambiguity attacks. This feature is summarized in Fig. \ref{Fig:CompareAA}.

\subsection{Security Against Watermark Error}
\label{Sect:CombinationAttack}
This section presents the establishment of the security boundary $r(n)$ for watermark detection errors, which serves as the threshold for ensuring the provable security of FedSOV. Specifically, if the watermark detection rate is greater than $r(n)$, FedSOV is deemed to be secure.

\subsubsection{Security Challenge }Federated model training and watermark removal attacks can bring errors to the detected watermark and cause a detection rate drop, which destroys the integrity of the compressed public keys \cite{li2022fedipr}. The verifier verifies the public key by extracting the watermark from the model.
In order to improve the robustness of the watermark, the verification process should tolerate certain errors between the extracted watermark and the original watermark.

However, an attacker can exploit this tolerance to construct a similarly forged watermark containing his public key to claim ownership. Specifically, the attacker tries to use his private key to generate a public key and forge the watermark, so that he can verify his ownership with his private key and forged watermark.

\begin{figure}
    \flushleft 
    \includegraphics[scale=0.4]{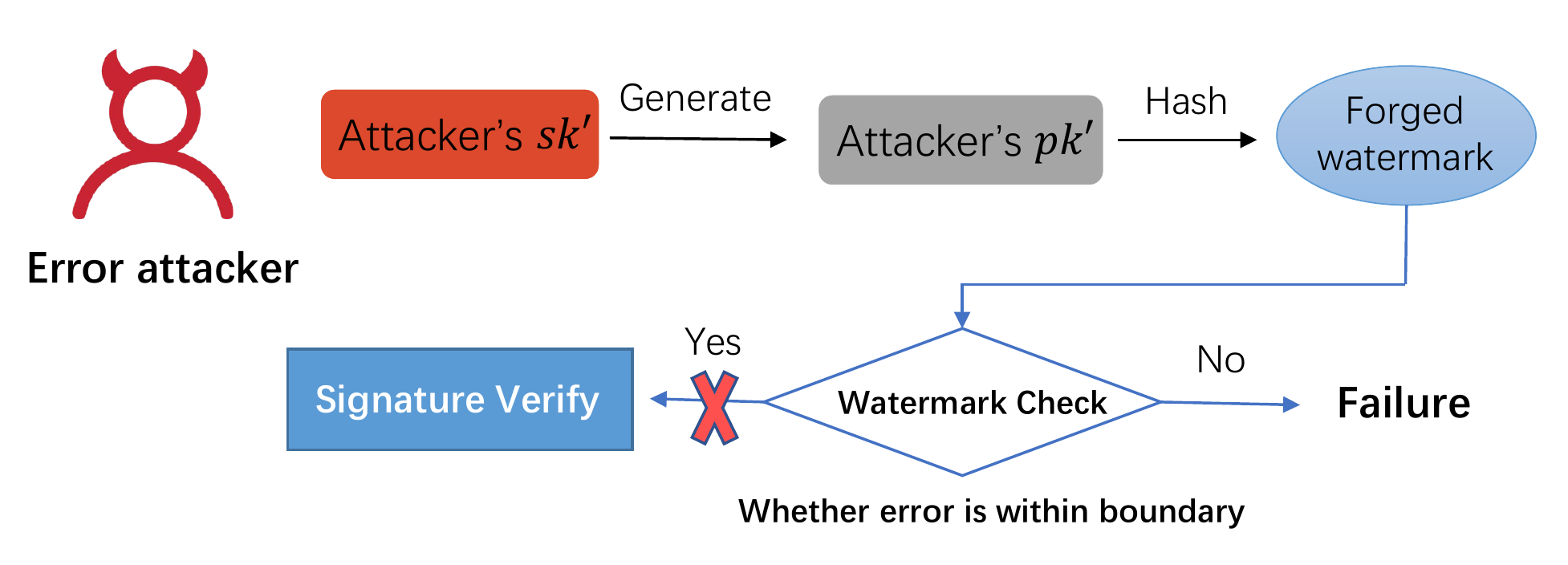}
    \caption{An error attacker tries to use his private key to generate a public key and forge the watermark, but FedSOV can resist it through a watermark check.}
    \label{Fig:WatermarkError}
\end{figure}



\subsubsection{Watermark Check }To address this security challenge, we propose \emph{security boundary of watermark detection rate} $r(n)$ defined in Def.\ref{Def:Security Boundary} to check whether the detected watermark is forged before signature verification. Fig. \ref{Fig:WatermarkError} is a simple schematic.
According to the analysis in Appendix \ref{Appe:SecurityBoundary}, \emph{when the watermark error is within this boundary range, the difficulty of finding hash near-collisions ensures that the probability of an attacker forging a valid watermark is negligible}. This boundary can be well satisfied as shown in experimental results Sect.\ref{Exp:EmbeddingEffect} and Sect.\ref{Exp:robustness}.
Therefore, FedSOV can provably overcome the security challenges based on the security of the hash function.



\begin{myDef}[\textbf{Watermark Detection Errors $err$}] \label{def:Watermark Detection Errors}
For a $n$ bits targeted hash watermark $\mathbf{h}$ and a given model $\mathbb {M}[\mathbf{W},\mathbf{h'}]$ with detected watermark $\mathbf{h'}$, the watermark detection rate is:
    \begin{equation}
        err =\rm{Dif} (\mathbf{h},\mathbf{h'})
    \end{equation}
    in which $\rm{Dif}(\mathbf{h},\mathbf{h'})$ is the Hamming distance between $\mathbf{h}$ and $\mathbf{h'}$.     
\end{myDef}

\begin{myDef}[\textbf{{Watermark Detection Rate $r$}}] \label{def:watermark-detect}
For a $n$ bits targeted hash watermark $\mathbf{h}$ and a given model $\mathbb {M}[\mathbf{W},\mathbf{h'}]$ with detected watermark $\mathbf{h'}$, the watermark detection rate is:
    \begin{equation}
        \label{watermark_det_rate}
        r=1-\frac{1}{n}err
    \end{equation}
    in which $err$ is the watermark detection errors in Def. \ref{def:Watermark Detection Errors}. 
\end{myDef}

\begin{myTheo}[\textbf{Security Boundary}]
\label{Def:Security Boundary}
Given the required low enough probability $P_A$ of an attacker forging a new valid watermark, the upper bound $err(n)$ of the watermark for $n$-bits detected watermark is uniquely determined by Eq. \eqref{Eq:UpperLimit}.
\begin{equation}
\label{Eq:UpperLimit}
    \begin{aligned}
   \frac{1}{2^n} \sum \limits_{i=0}\limits^{ 2err(n)-1 } C^i_n< P_A&\leq\frac{1}{2^n} \sum \limits_{i=0}\limits^{ 2err(n) } C^i_n.
    \end{aligned}
\end{equation}

To facilitate subsequent experimental verification for $err(n)$, the \textbf{lower bound} $r(n)$ of watermark detection rate is computed as:
\begin{equation} \label{eq:lower-bound-r}
    r(n) = 1-err(n) /n,
\end{equation}
\end{myTheo}

When the watermark detection rate is above the $r(n)$ (or the watermark error falls within the $err(n)$, the probability of an attacker forging a valid watermark is negligible due to the difficulty of finding hash near-collisions. As evidenced by our experimental results in Sect. \ref{Exp:EmbeddingEffect} and Sect. \ref{Exp:robustness}, this boundary can be satisfactorily met.  The secure analysis and proof of Theorem 3 and the specific watermark check protocol are shown in Appendix \ref{Appe:SecurityBoundary}.

\textbf{Remark:} As an illustration of the security boundary $r(n)$, let us consider a 2048-bit watermark. If the detection rate is greater than the security boundary value $r(2048)$ = 82.1\%, the probability of an attacker forging a valid watermark is estimated to be less than $1/2^128$. This probability is deemed negligible, and hence, the risk of successful forgery can be disregarded.
\section{Experimental Analysis}
\label{Sect:Experimental Analysis}

This section illustrates the empirical study of federated DNN models with FedSOV watermarks. We use classic models of computer vision and natural language processing to analyze FedSOV from the perspectives of \textit{fidelity, reliability, robustness, and security of watermarks}. Besides, the time cost of ownership verification can be seen in Appendix \ref{Exp:TimeCost}.

\subsection{Datasets and DNN Models}

For the image classification task, we choose the classic AlexNet \cite{krizhevsky2017imagenet} and ResNet18 \cite{he2016deep} models with datasets CIFAR10 and CIFAR100 \cite{krizhevsky2009learning}. For natural language tasks, we choose the DistilBERT model \cite{sanh2019distilbert} with datasets SST2 and QNLI. The detailed hyper-parameter settings and model structures are shown in the Appendix \ref{Sect:AppendixExperiment}

\subsection{Evaluation Metrics}
We evaluate our ownership verification scheme from five perspectives: Fidelity, Reliability, Robustness, Security, and Time Cost.

\textbf{Fidelity:}
A well-behaved FMOV scheme should have as little impact on the main task performance of the model as possible, so we use the model main accuracy $Acc_{main}$ as the evaluation index of Fidelity.

\textbf{Reliability: }
A well-behaved FMOV scheme should be able to embed the watermark effectively in various scenarios, including various numbers of clients and various watermark lengths. We choose the \textit{watermark detection rate} $r$ (see Def. \ref{def:watermark-detect}) as the main indicator to analyze the reliability of FedSOV under various scenarios.

\textbf {Robustness:} A well-behaved FMOV scheme should be able to resist various \textit{watermark removal attack} and be preserved as much as possible after a series of model parameter changes, so we choose the \textit{watermark detection rate} $r$ as the robustness evaluation index.

\textbf{Security: }
This evaluation is set to demonstrate that the security boundary for watermark check in Sect.\ref{Sect:CombinationAttack} is reasonable and can be well satisfied.
When analyzing the reliability in various federated training scenarios and the robustness of watermarking under various attacks, we should also  use $r(n)$ as the main metric to measure security to ensure that FedSOV can provide a secure FMOV method against secure challenge caused by watermark error.
The security boundary is verified in Sect. \ref{Exp:EmbeddingEffect} and Sect. \ref{Exp:robustness}.


\subsection{Fidelity}
To evaluate the fidelity of FedSOV watermarking, we compare  the $Acc_{main}$ of the FedSOV model embedded with watermark and the $Acc_{main}$ of the classic FedAvg model without an embedded watermark. 
We use the AlexNet model on the Cifar10 dataset, the ResNet18 model on the Cifar100 dataset and the DistilBert on SST2 and QNLI datasets to test the two schemes with different numbers of clients.

\begin{figure}[htbp]
    
    \centering
    \subfigure[ResNet18 with CIFAR10]{
        \begin{minipage}[t]{0.5\linewidth}
        \centering
        \includegraphics[width=1.7in]{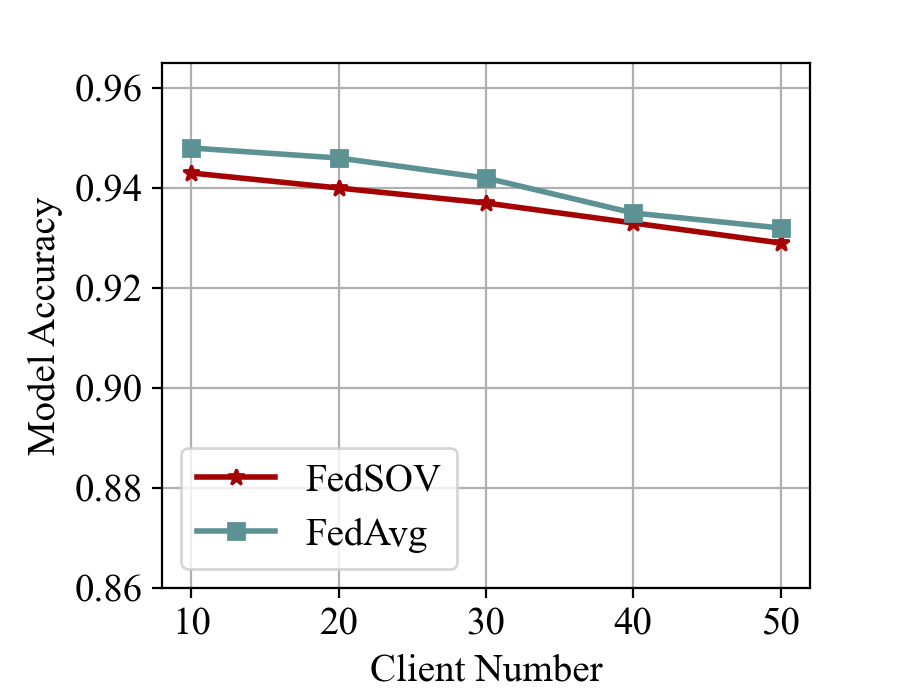}
        \end{minipage}%
    }%
    \subfigure[AlexNet with CIFAR100]{
        \begin{minipage}[t]{0.50\linewidth}
        \centering
        \includegraphics[width=1.7in]{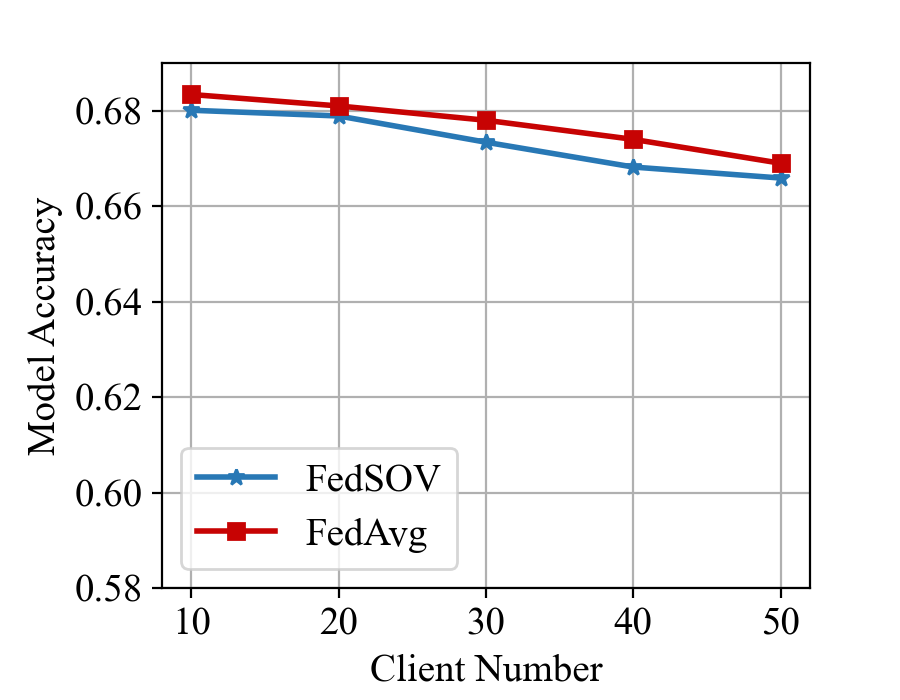}
        \end{minipage}%
    }%

    \subfigure[DistilBert with SST2]{
        \begin{minipage}[t]{0.5\linewidth}
        \centering
        \includegraphics[width=1.7in]{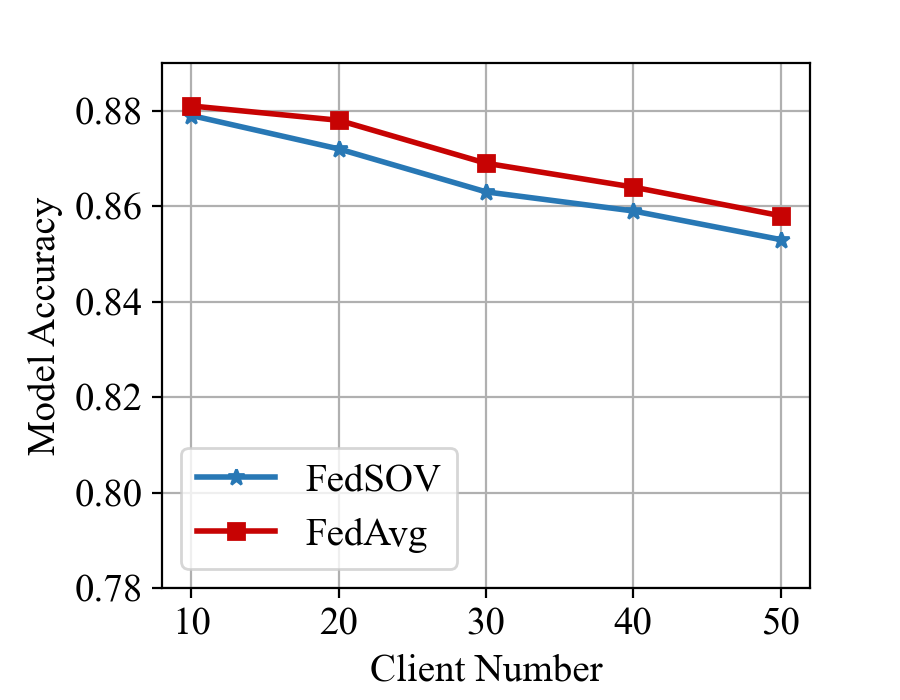}
        \end{minipage}%
    }%
    \subfigure[DistilBert with QNLI]{
        \begin{minipage}[t]{0.50\linewidth}
        \centering
        \includegraphics[width=1.7in]{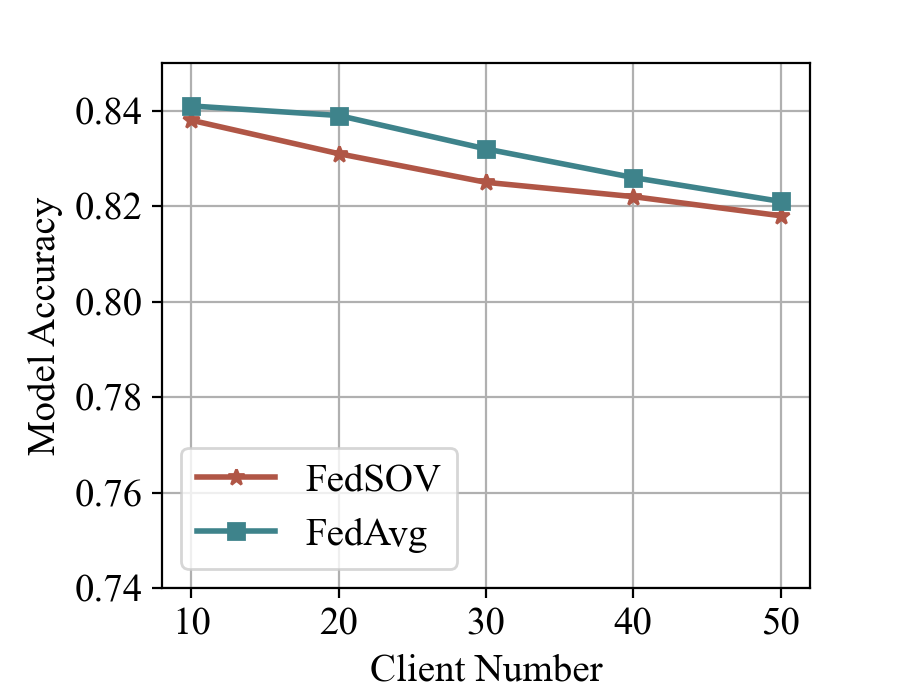}
        \end{minipage}%
    }%
    \centering
    \caption{Figure (a)-(d) illustrates the main accuracy of image and text classification models with various numbers of clients.} 
    \label{Fig:Fidelity}

\end{figure}

\begin{figure*}
    \centering
    \subfigure[ResNet18 with CIFAR100]{
        \includegraphics[scale=0.38]{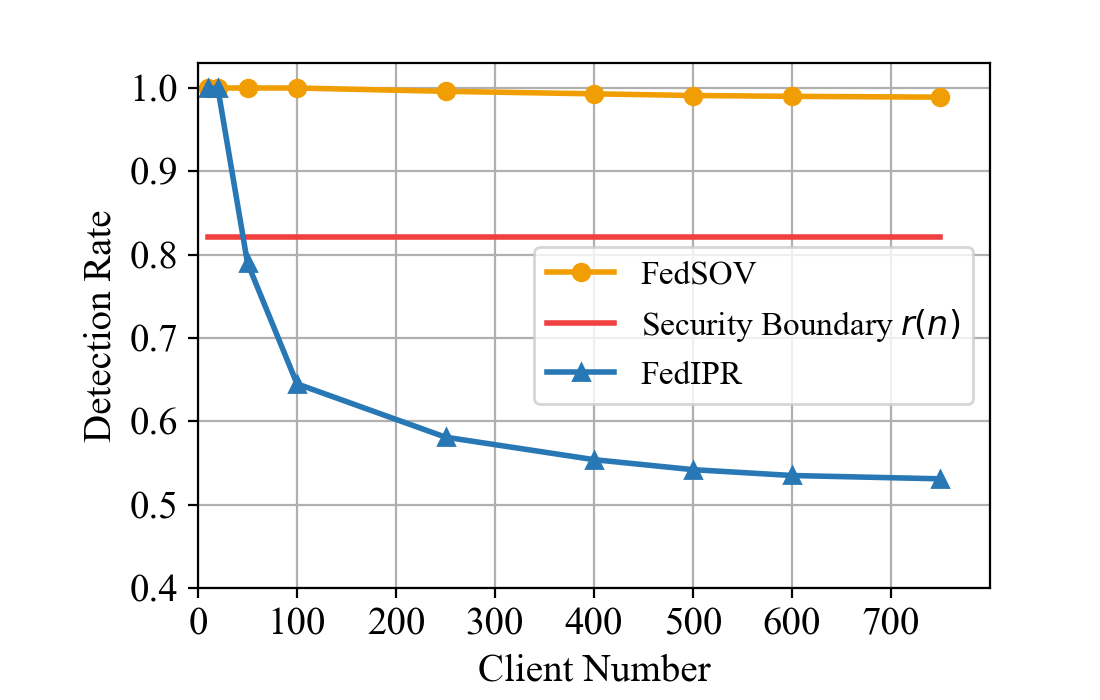} 
    }
    \subfigure[AlexNet with CIFAR10]{
        \includegraphics[scale=0.38]{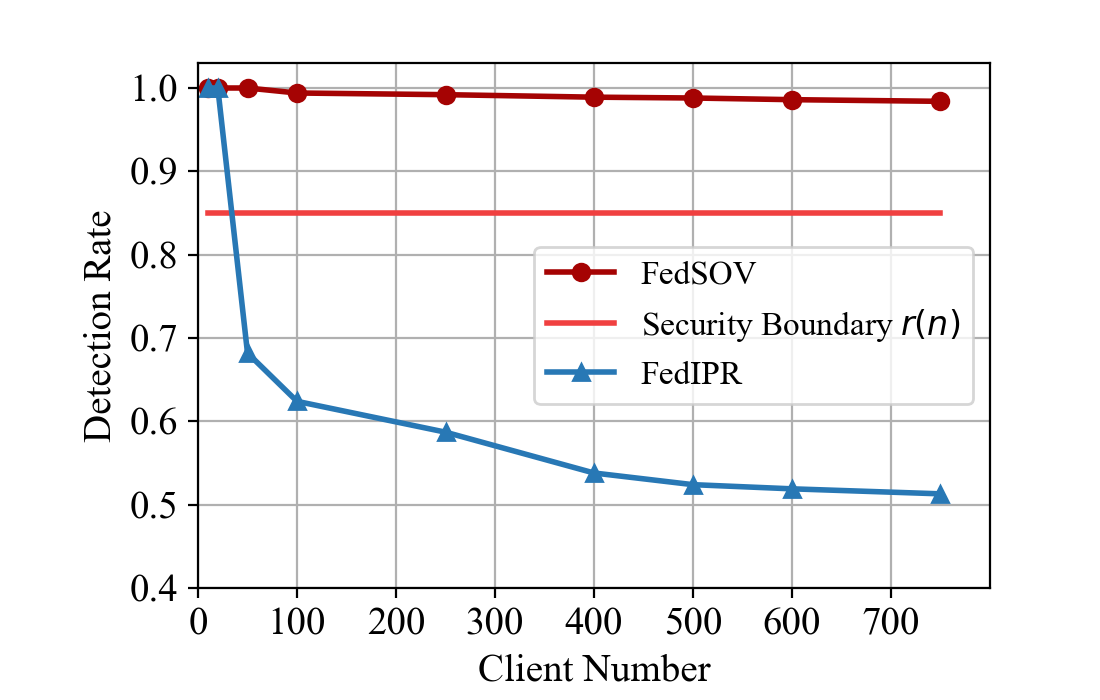} 
    }
    \subfigure[DistilBert with SST2]{
        \includegraphics[scale=0.38]{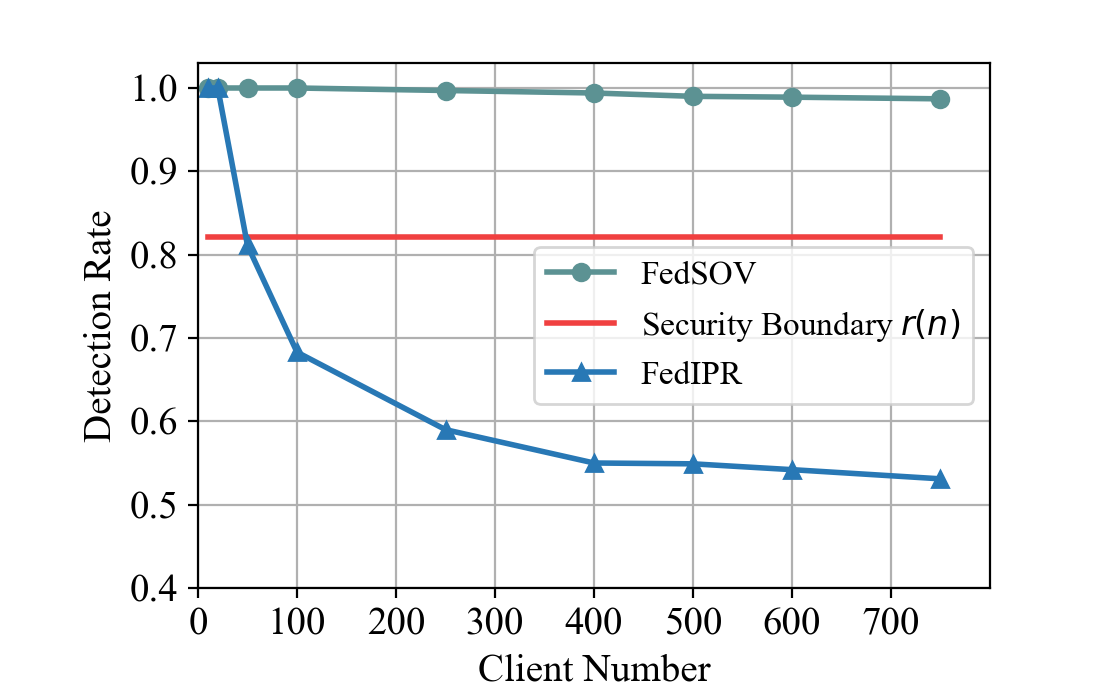}
    }
    \DeclareGraphicsExtensions.
    \caption{Figure (a)-(c) demonstrate the watermark detection rate in image and text classification tasks with varying client numbers, in which ResNet and DistilBert have 2048 bits watermark and AlexNet has 1024 bits watermark.}
    \label{Fig:ClientNum}
\end{figure*}

\begin{figure*}
    \centering
    \subfigure[ResNet18 with CIFAR100]{
        \includegraphics[scale=0.38]{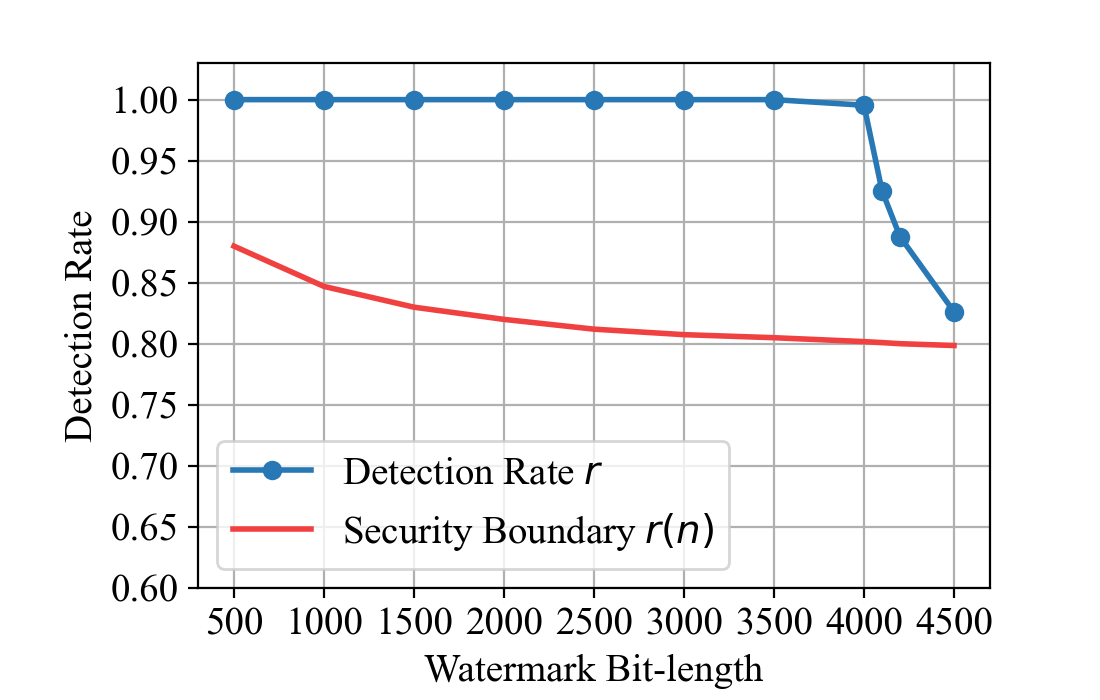} 
    }
    \subfigure[AlexNet with CIFAR10]{
        \includegraphics[scale=0.38]{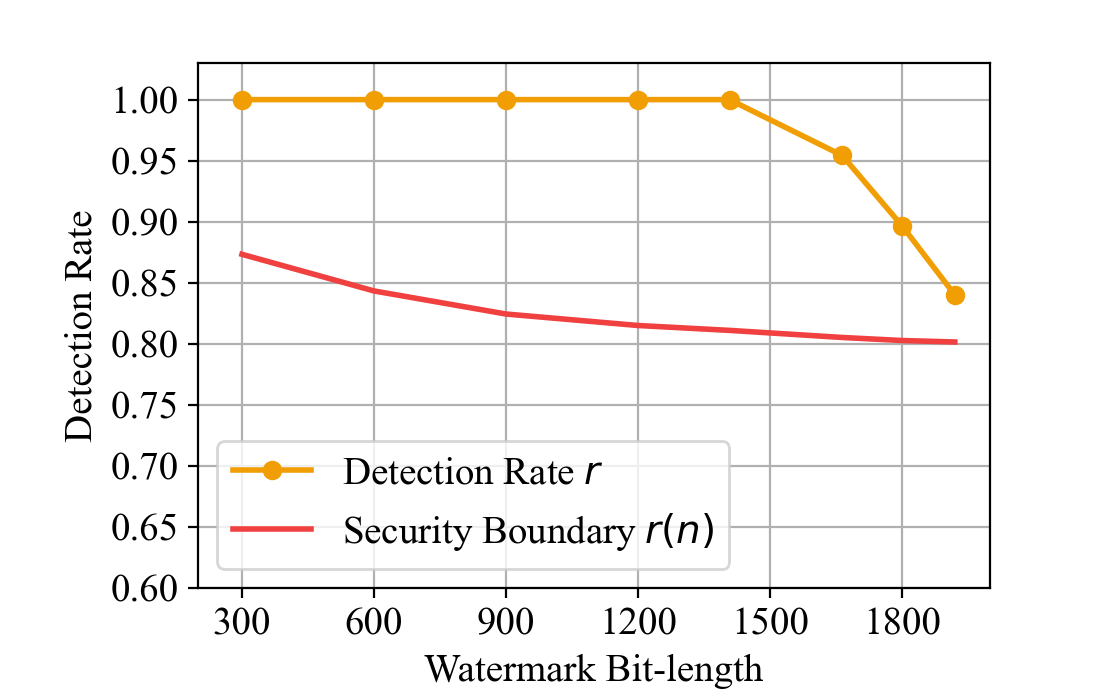} 
    }
    \subfigure[DistilBert with SST2]{
        \includegraphics[scale=0.38]{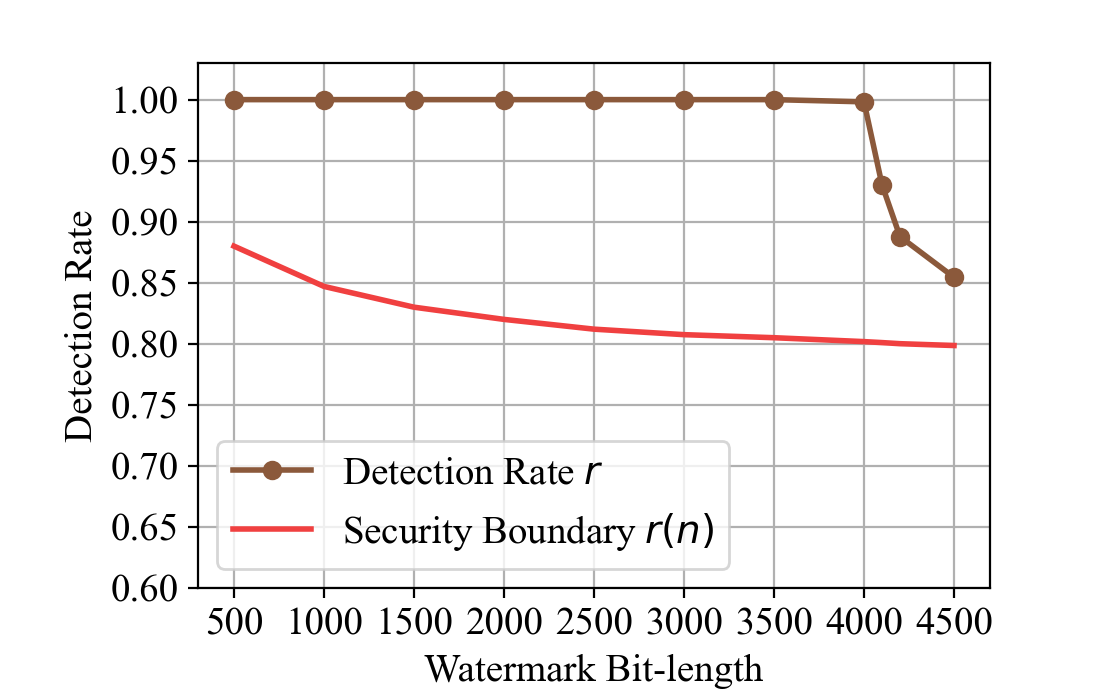}
    }
    \DeclareGraphicsExtensions.
    \caption{Figure (a)-(c) show the detection rate and secure boundary with the  change of watermark length of image and text classification models with 200 clients.}
    \label{Fig:WmLength}
\end{figure*}

Fig. \ref{Fig:Fidelity} shows the main task accuracy of FedSOV with minor performance drops of less than 2\% compared to FedAvg. This may be due to the regularization of embedding watermarks that limits the optimization space of model parameters.

\subsection{Reliability}
\label{Exp:EmbeddingEffect}
The task of embedding a watermark with a length of $n$ is mainly completed by each client through training and updating, and the embedded watermark should be detected as completely as possible, so as to be used for ownership verification. The watermark embedding effect is affected by \emph{the number of clients and the length of the watermark}. Moreover, the detection rate $r$ should be above the security boundary provided in Theo. \ref{Def:Security Boundary}.

\textbf{Client Number: }
One of the innovations of FedSOV is that compared with previous solutions, it can support a federated learning system with a large number of clients. Here we demonstrate this advantage by comparing it with FedIPR \cite{li2022fedipr} 100-bit-length credential information per client. 

Fig. \ref{Fig:ClientNum} shows the watermark detection rate of FedIPR drops considerably as the number of clients increases, eventually converging to approximately 50\%. In contrast, FedSOV maintains a watermark detection rate of over 99\%, with a minimal drop of less than 1\%, demonstrating that the number of clients has a negligible effect on watermark embedding. Moreover, the watermark detection rate $r$ of FedSOV consistently exceeds the security boundary $r(1024)=0.8505$ and $r(2048)=0.8217$, thereby satisfying the requisite security standards.

\textbf{Watermark Length: }
We study the embedding effect and security boundary of watermarks with different lengths on different models, so as to provide guidance for choosing the appropriate watermark length when using FedSOV.

From Fig. \ref{Fig:WmLength},  we can obtain two observations: a) with the increase of the watermark length, the watermark detection rate is initially maintained at almost 100\%, and when it exceeds a certain threshold, it drops rapidly; b) as the watermark length increases, the security boundary $r(n)$ decreases more and more slowly and tends to converge.
Although FedSOV is compatible with a wide range of watermark lengths, to enhance the resilience of watermarks, a watermark length should be chosen that ensures a relatively high watermark detection rate.

\subsection{Robustness}
\label{Exp:robustness}

In this subsection, we demonstrate the robustness of FedSOV watermarking against post-processing watermark removal techniques.

\begin{figure}[htbp]
    \centering
    \subfigure[Fine-tuning Attack]{
        \label{Fig:FT-attack}
        \begin{minipage}[t]{0.5\linewidth}
        \centering
        \includegraphics[width=1.7in]{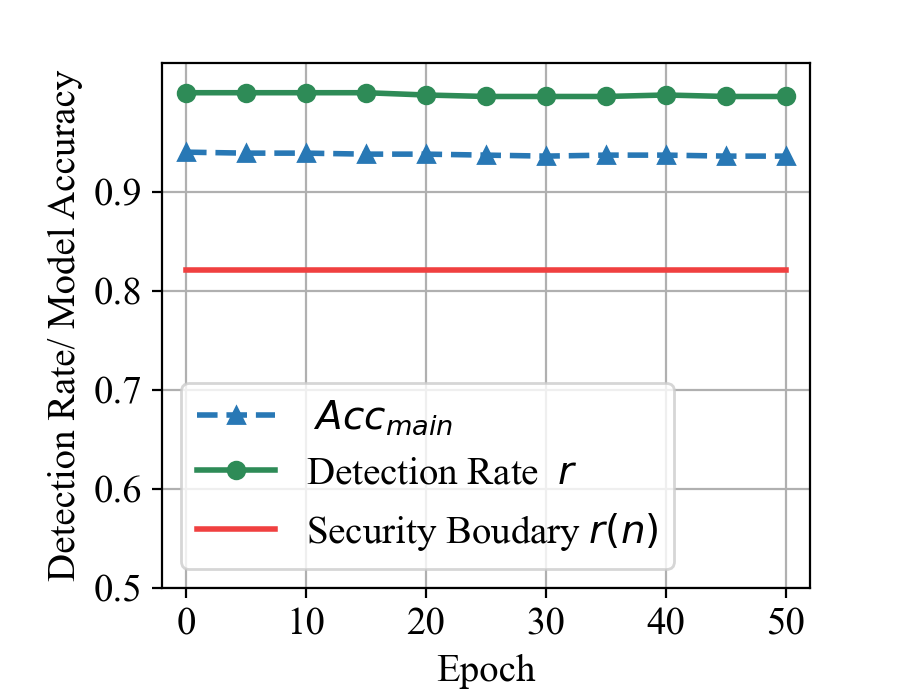}
        \end{minipage}%
    }%
    \subfigure[Pruning Attack]{
        \label{Fig:Purning-atttack}
        \begin{minipage}[t]{0.50\linewidth}
        \centering
        \includegraphics[width=1.7in]{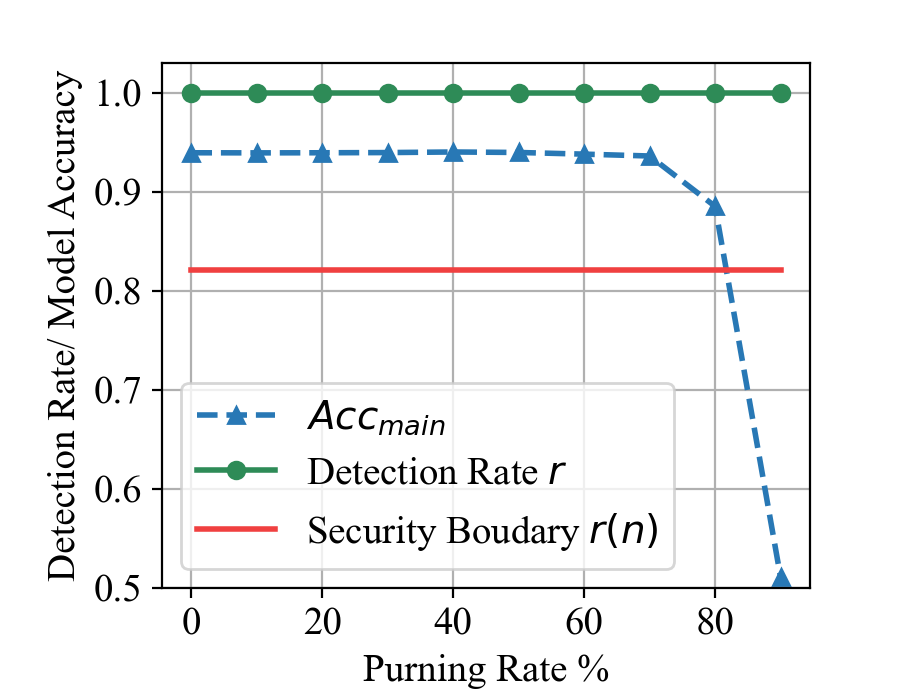}
        \end{minipage}%
    }%
    \centering
    \caption{Figure (a)-(b) describe the robustness against fine-tuning and pruning attack of ResNet18 in CIFAR10 }
\end{figure}

Fine-tuning and pruning are general watermark removal techniques that do not require additional model and data knowledge. Their goal is to remove or destroy the watermark contained in the model parameters while preserving the performance of the model as much as possible, making the watermark detection incomplete and insufficient for ownership verification. 

Besides, since the FedSOV watermark information including embedding matrix and position is public, the attacker can use the public information to carry out the \emph{targeted destruction attack} after stealing the model. The experimental results and analysis of robustness against targeted destruction attack is shown in Appendix \ref{Sect:AppendixTarget}.

\subsubsection{Robustness against Fine-tuning Attack}
In this experiment, we conducted fine-tuning by removing the regularization term $L_h$ from the main loss. The result of the experiment is shown in Fig. \ref{Fig:FT-attack}, which indicates that even after 50 Epochs, the detection rate of the 2048-bit watermark in ResNet18 using FedSOV remains almost 100\%, above the security boundary.
\subsubsection{Robustness against Pruning Attack}
We adopted the pruning scheme in \cite{see2016compression} to remove redundant parameters of the model. We evaluate model robustness using detection rate and master task accuracy at different pruning rates. Fig. \ref{Fig:Purning-atttack} shows that the 2048 bits watermark of FedSOV in ResNet18 always has a detection rate close to 100\%.

\section{Conclusion}
\label{Sect:Conclusion}
To allow model watermarking to support a large number of clients in FL systems, and resist the ambiguity attack, we propose a scheme for ownership verification based on digital signature. The experiment shows that the number of clients has almost no effect on the effectiveness of our embedded watermark, which allows our solution to be applied in FL scenarios with a large number of clients. Security analysis shows that it is infeasible for an attacker to successfully launch the ambiguity attack under the unforgeable digital signature. What's more, we overcome the challenge of the watermark error using near-collision resistance hash. Security analysis and experiment results show that the verifier can also correctly verify ownership with overwhelming probability when the watermark detection rate is greater than the security boundary.

{\small
\bibliographystyle{unsrt}
\bibliography{reference}
}

\appendix
\section{Appendix}

\subsection{Analysis of Security against Watermark Error}
\label{Appe:SecurityBoundary}
\subsubsection{Specific watermark check protocol}
Instead of using the concatenated public key $pk_{con}$ directly as the watermark during watermark generation, FedSOV uses the hash value of $pk_{con}$ as the watermark to protect the public key. 

The following two steps are needed to check the watermark before signature verification: 
\begin{enumerate}
    \item The client $c_i$ needs to let the server send $pk_{con}$ to the verifier.
    \item The verifier computes the Hamming distance between the $\rm {H}(pk_{con})$ and the watermark $\mathbf{h}'$, if the distance is less that $err(n)$, the verifier accepts the $pk_{con}$, else he believes that $pk_{con}$ is incorrect.
\end{enumerate}

\subsubsection{Proof of Theorem 3}

For the stolen model, the attacker can first modify $err(n)$ bits of $\mathbf h$ into the damaged watermark $\mathbf h'$, and then find the $err(n)$-near-collision for $\mathbf h'$, so that he only needs to find $2err(n)$-near-collision for the original watermark to successfully replace the public key in the watermark. In this way, an attacker can find a set of public-private key pairs, and the hash of the public key can pass the near-collision verification. 

We show that the probability of an attacker replacing the public key in the watermark is negligible when the upper limit $err(n)$ of the attacker's damaged watermark bits is within a certain limit. We denote $C_y^x$ as the combination number, where
\begin{equation}
    C_y^x=\prod\limits_{i=y-x}^yi/\prod\limits_{j=1}^xj.
\end{equation}

\begin{myTheo}
\label{Theo:NearCollision}
Assume the output of hash function $\rm{H}(\cdot):\mathbb G\rightarrow\{0,1\}^{n}$ is random \cite{ro:bellare1993random}. The attacker gets $q$ watermarks and computes hash function $k$ times. The upper limit of the attacker's damaged watermark bits is $err(n)$. The probability $P_A$ that the attacker successfully replaces the public key in any of the $q$ watermarks satisfies:

\begin{equation}
    \begin{aligned}
    P_A\leq\frac{kq}{2^n}\cdot\sum^{2err(n)}_{i=0}C_n^i.
    \end{aligned}
\end{equation}
\end{myTheo}

\noindent \textbf{Proof}.

After the attacker $\mathcal{A}$ steals $q$ models and gets $q$ watermarks $\mathbf{h}_1,...,\mathbf{h}_q\in\{0,1\}^{n}$, he will generate a public key $pk$ and compute its hash value $\mathbf{h}$ by $\rm{H}(\cdot)$. The probability $P_1$ that a $2err(n)$-near-collision doesn't happen between the hash function output and the $q$ watermarks in a single computation satisfies

\begin{equation}
\label{Equa:SingleNearCollision}
    \begin{aligned}
        P_1&\geq1-q\cdot\frac{\sum^{2err(n)}_{i=0}C_n^i}{2^{-n}}.
    \end{aligned}
\end{equation}

$2^{n}$ is the size of the output space of the hash function $\rm H(\cdot)$, $q\sum^{2err(n)}_{i=0}C_n^i$ is the size of the space that $2err(n)$-near collision happens between the hash function output and the $q$ watermarks. Because the output of the hash function is indistinguishable from a random number \cite{ro:bellare1993random}, the probability that the output of the hash function falls on all points in $\{0,1\}^{n}$ is equal. So, we can express the $2err(n)$-near-collision probability by dividing the size of the $2err(n)$-near-collision space by the total size of the hash output space (i.e., Eq. \eqref{Equa:SingleNearCollision}).

When $\mathcal{A}$ computes $\rm H(\cdot)$ $k$ times, the probability $P_A$ that the $2err$-near-collision doesn't happens satisfies

\begin{equation}
\label{Equa:KTimesCollision}
    \begin{aligned}
        P_A&\leq 1-(1-q\cdot2^{-n}\cdot\sum^{2err(n)}_{i=0}C_n^i)^k\\
        &=1-\sum^{k}_{i=0}C_k^i(-q\cdot2^{-n}\cdot\sum^{2err(n)}_{j=0}C_n^j)^i\\
        &\leq kq\cdot2^{-n}\cdot\sum^{2err(n)}_{i=0}C_n^i.
    \end{aligned}
\end{equation}

\begin{figure}
    \centering
    \includegraphics[scale=0.15]{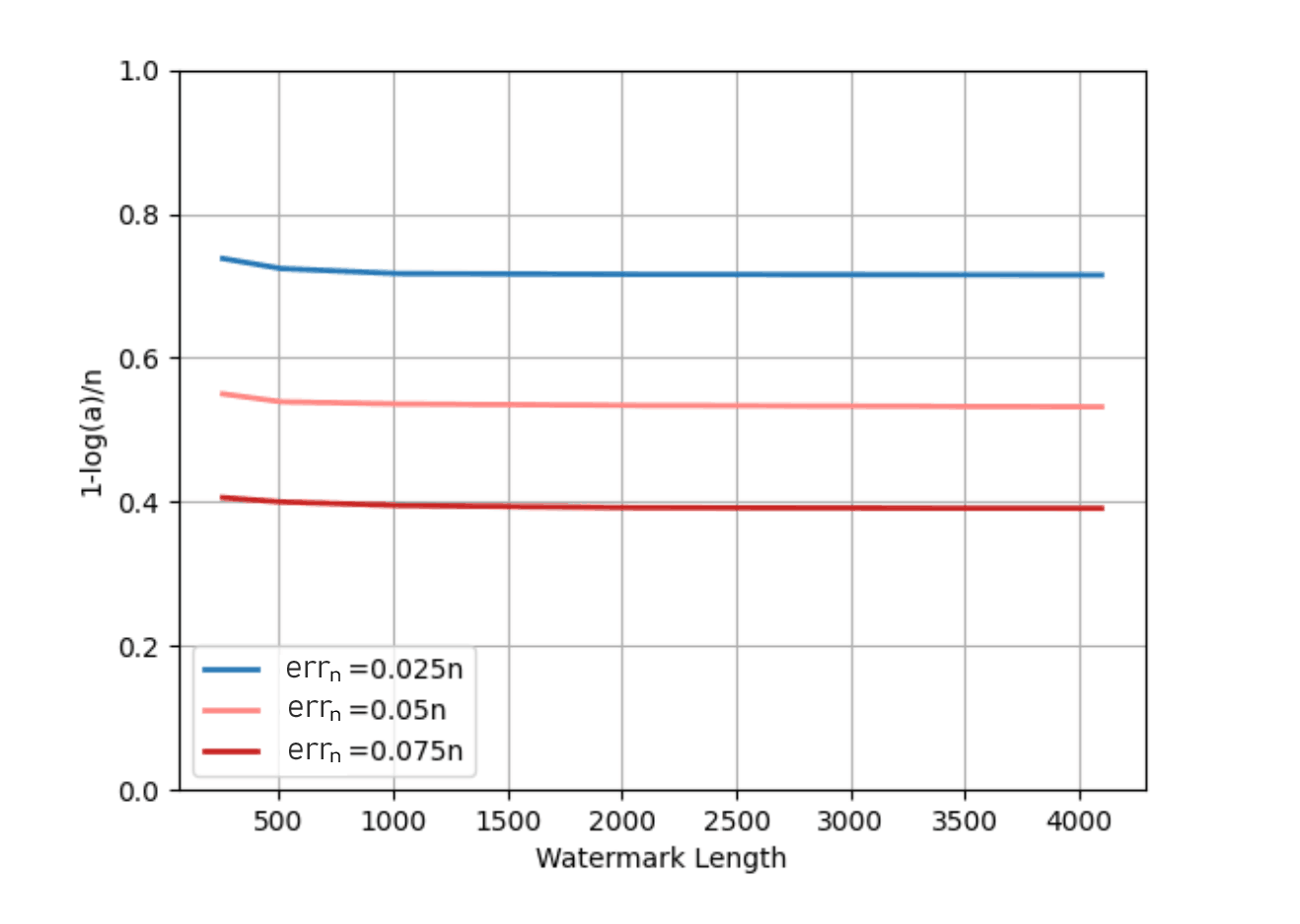}
    \caption{This figure illustrates that the value of $1-log(a)/n$ converges as $n$ increases}
        \label{Fig:log}
\end{figure}

So the Theo. \ref{Theo:NearCollision} holds. We will show that $P_A$ is small enough to be negligible.

Let $a=\sum^{2err(n)}_{i=0}C_n^i$. As Fig. \ref{Fig:log} illustrates, the value of $1-log(a)/n$ tends to converge as the watermark length $n$ grows for the same $err(n)$. For the same watermark length, the value of $log(a)/n$ decreases with increasing $err$. When $err(n)=\lfloor0.025n\rfloor$, $log(a)/n$ converges to 0.715, $P_A\leq kq\cdot\frac{1}{2^{0.715n}}$. When $err=\lfloor0.075n\rfloor$, $log(a)/n$ converges to 0.39, $P_A\leq kq\cdot\frac{1}{2^{0.39n}}$.

Since $k,q$ are polynomial sizes which are negligible compared with $2^{0.713n}$ and $2^{0.39n}$, the $P_A$ is negligible. which proves that it's hard for the attacker to corrupt the $pk_a$ in the watermark.

\subsection{Proof of Theorem 1}
\label{Apdix:Proof1}
\begin{myTheo}
\label{Theo:Appendix-AA}
For a targeted forged watermark $B'=(b'_1,\cdots,b'_n)$ and targeted model parameters $\mathbf {W}_t=(w_1,w_2,\cdots,w_{\omega})$, it is easy to  find a forged embedding matrix $\mathbf E'=\{e_{ij}\}_{\omega \times n}$ for FedIPR that staisties $\rm sgn (\mathbf {W}_t \mathbf {E'} )=B'$, i.e.,

\begin{equation}
\left\{\begin{array}{c}    (w_1e_{11}+w_2e_{21}+\cdots+w_{\omega}e_{\omega 1})b'_1>0 \\    (w_1e_{12}+w_2e_{22}+\cdots+w_{\omega}e_{\omega 2})b'_2>0 \\    \cdots \\     (w_1e_{1n}+w_2e_{2n}+\cdots+w_{\omega}e_{\omega n})b'_n>0\end{array}\right. .
\end{equation}
The three conditions in Def.\ref{Def:Ambiguity Attack} are all satisfied. An ambiguity attack can be successfully constituted for FedIPR models.
\end{myTheo}

\emph{Proof.} For the convenience of explanation, we consider the sufficient condition of the above theorem: if finding a $\mathbf E'=\{e_{ij}\}_{\omega \times n}$  that satisfies $ \mathbf {W}_t \mathbf {E'} =B'$, i.e.,
\begin{equation}
\left\{\begin{array}{c}    w_1e_{11}+w_2e_{21}+\cdots+w_{\omega}e_{\omega 1}=b'_1 \\    w_1e_{12}+w_2e_{22}+\cdots+w_{\omega}e_{\omega 2}=b'_2 \\    \cdots \\     w_1e_{1n}+w_2e_{2n}+\cdots+w_{\omega}e_{\omega n}=b'_n\end{array}\right.
\label{Equ:SuffientCon}
\end{equation}
is easy, forging a matrix $\mathbf E'$ that staisties $\rm sgn (\mathbf {W}_t \mathbf {E'} )=B'$ is easy.

Each row of equations in Eq. \eqref{Equ:SuffientCon} is independent. Since the rank of the augmented matrix for each linear equation is less than $\omega$, i.e., $ rank (\mathbf {W}_{t},b'_i)=1 < \omega$, an infinite number of solutions can be found for each equation. Therefore, an attacker can easily forge an infinite number of embedding matrices $\mathbf E'$ that satisfies $ \mathbf {W}_t \mathbf {E'} =B'$, i.e., the condition (1) and (2) in Def. \ref{Def:Ambiguity Attack} hold.
Besides, the above forging process doesn't change the model parameters, so the inference performance of the model is not degraded, i.e., $|\mathcal F(\mathbb M[\mathbf W,\mathbf B']), \mathbf {D}_t)-\mathcal {F}_t|=0 \leq \epsilon _f$. Thus, the condition (3) of ambiguity attack holds. In summary, an ambiguity attack can be successfully constituted for FedIPR models.

\noindent\textbf{Example:} For parameters $\mathbf {W}_t=(0.1048,  0.1316,  0.1148, -0.1190)$, watermark $\mathbf B'=(1,\ 1,\ -1,\ 1)$, it is easy to find $\mathbf E'=$ 
$$\left (\begin{array}{cccc}
6.16055 &-0.5&-0.5&-4 \\
7.26999&-0.3&-0.7&-3  \\
-10.80407&0.1&0.8&0.2 \\
8.08188&-0.4&-0.5&-2.2\\
\end{array}\right)$$ 
that staisties $\rm sgn (\mathbf {W}_t \mathbf {E'} )=B'$. Therefore, an ambiguity attacker can claim his ownership with $\mathbf E'$ and $\mathbf B'$.

\begin{table*}[]
\centering

\caption{Training parameters for Federated AlexNet, ResNet18 and DistilBERT}
\begin{tabular}{c|c|c|c}
\hline
Hyper-parameter     & AlexNet                   & ResNet-18                  & DistilBERT                  \\ \hline
Optimization method & SGD                       & SGD                        & SGD                         \\
Learning rate       & 0.01                      & 0.01                       & 0.01                        \\
Batch size          & 16                        & 16                         & 32                          \\
Global Epochs       & 200                       & 200                        & 80                          \\
Local Epochs        & 2                         & 2                          & 1                           \\
Learning rate decay & 0.99 at each global Epoch & 0.100 at each global Epoch & linear schedule with warmup \\
Regularization Term & BCE loss, Hinge-like loss & BCE loss, Hinge-like loss  & BCE loss, Hinge-like loss   \\ \hline

\end{tabular}
\label{Table:Traning Parameters}
\end{table*}

\begin{table}[]
\caption {AlexNet Architecture used in FedSOV (Embed feature-based
watermarks across Conv3, Conv4 and Conv5)}
\begin{tabular}{c|c|c|c}
\hline
Layer name & Output size  & Weight shape                     & Padding \\ \hline
Conv1      & 32 $\times$ 32 & 64 $\times$ 3$\times$ 5$\times$ 5      & 2       \\
MaxPool2d  & 16 $\times$ 16 & 2 $\times$ 2                       &         \\
Conv2      & 16 $\times$ 16 & 192 $\times$ 64 $\times$ 5 $\times$ 5  & 2       \\
Maxpool2d  & 8 $\times$ 8   & 2 $\times$ 2                       &         \\
Conv3      & 8 $\times$ 8   & 384 $\times$ 192 $\times$ 3 $\times$ 3 & 1       \\
Conv4      & 8 $\times$ 8   & 256$\times$ 384 $\times$ 3 $\times$ 3  & 1       \\
Conv5      & 8 $\times$ 8   & 256 $\times$ 256 $\times$ 3 $\times$ 3 &         \\
MaxPool2d  & 4 $\times$ 4   & 2 $\times$ 2                       &         \\
Linear     & 10           & 10 $\times$ 4096                   &         \\ \hline
\end{tabular}
\label{Table:Alex}
\end{table}

\begin{table}[]
\caption {ResNet18 Architecture used in FedSOV (Embed
watermarks across Res5 Block)}
\begin{tabular}{c|c|c|c}
\hline
Layer name & Output size  & Weight shape                     & Padding \\ \hline
Conv1      & 32 $\times$32 & 64 $\times$3$\times$5$\times$5      & 1       \\ \hline
Res2       & 32 $\times$32 & $\left [ \begin{matrix}
64 \times 64 \times 3 \times 3 \\
64\times64\times3\times3 \\
\end{matrix} \right ]$ $\times$2                    & 1       \\ \hline
Res3       & 16 $\times$16 & $\left [ \begin{matrix}
128\times128\times3\times3 \\
128\times128\times3\times3 \\
\end{matrix} \right ] $ $\times$2  & 1       \\ \hline
Res4       & 8 $\times$8   & $\left [ \begin{matrix}
256\times256\times3\times3 \\
256\times256\times3\times3 \\
\end{matrix} \right ]$ $\times$2                   & 1       \\ \hline
Res5       & 4 $\times$4   & $\left [ \begin{matrix}
512\times512\times3\times3 \\
512\times512\times3\times3 \\
\end{matrix} \right ]$ $\times$2 & 1       \\ \hline
Linear     & 100          & 100 $\times$512                   &         \\ \hline
\end{tabular}
\label{Table:ResNet18}
\end{table}

\begin{figure*}
    \centering
    \subfigure[ResNet18 with 1024 bits Watermark]{
        \label{Target_Res_1024}
        \includegraphics[scale=0.38]{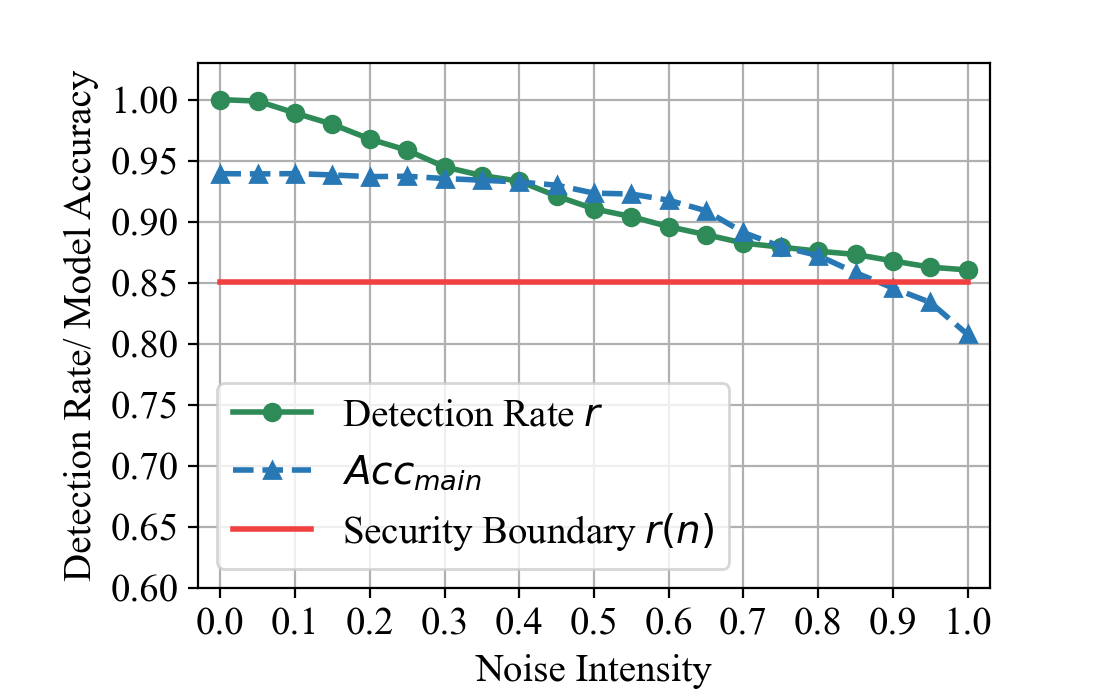} 
    }
    \subfigure[ResNet18 with 2048 bits Watermark]{
        \label{Target_Res_2048}
        \includegraphics[scale=0.38]{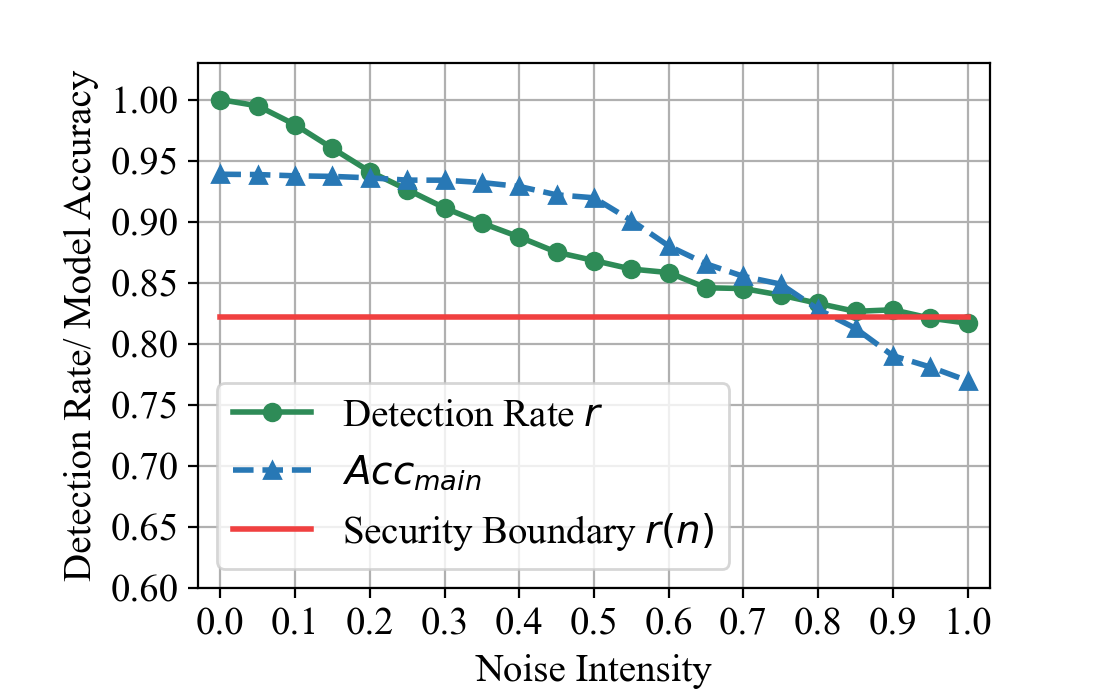} 
    }
    \subfigure[DistilBert with 3076 bits Watermark]{
        \label{Target_DistilBert_3076}
        \includegraphics[scale=0.38]{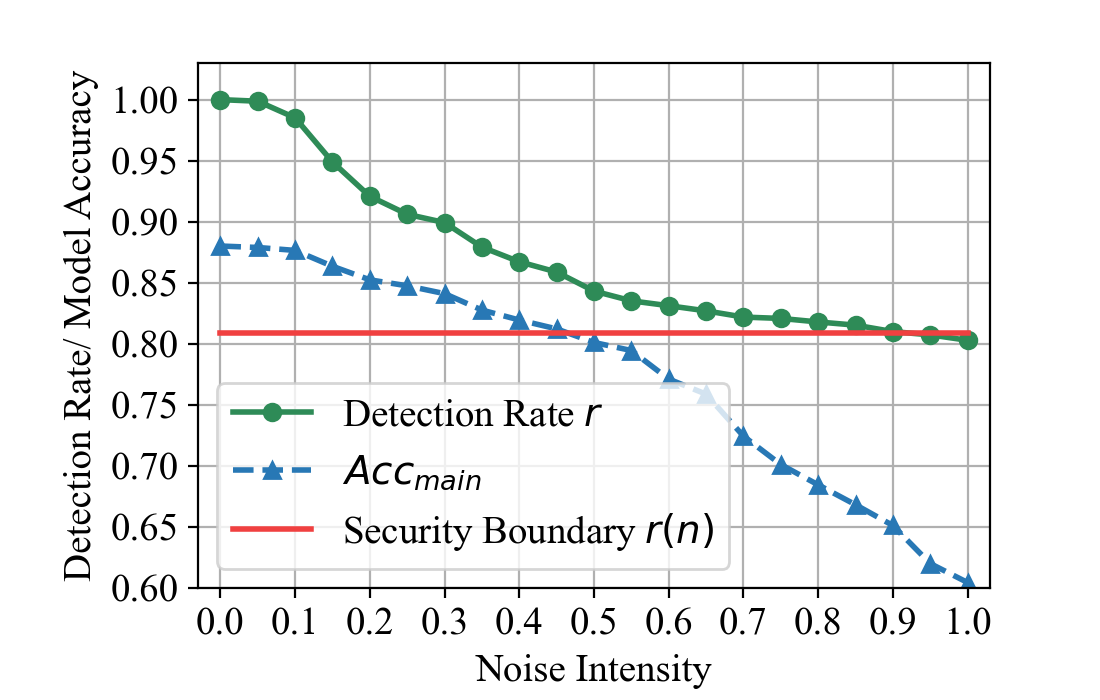}
    }
    \DeclareGraphicsExtensions.
    \caption{Figure (a)-(c) show the robustness to targeted destruction attack of ResNet18 on CIFAR10 and DistilBert on SST2.}
    \label{Fig:TargetedAttack}
\end{figure*}
\subsection{Short Signature Scheme}
\label{Appe:ShortSignature}
\begin{myDef}[\textbf{Bilinear Map}]
The bilinear map is sometimes used to construct digital signature schemes. The bilinear system \cite{signature:boneh2001weil} is defined as $\mathbb{S}=(p, \mathbb{G}, \mathbb{G}_T, e)$, where $p$ is the order of the multiplicative cyclic groups $\mathbb{G}, \mathbb{G}_T$. $e$ is a bilinear map, $e:\mathbb{G}\times\mathbb{G}\rightarrow\mathbb{G}_T$.  which has three properties:
\begin{enumerate}
    \item Computability: For any $g,h\in\mathbb{G}$, $e(g,h)$ can be computed effciently.
    \item Bilinear:  For any $g,h\in\mathbb{G}$, $a,b\in\mathbb{Z}_p$, it has $e(g^{a},h^{b})=e(g,h)^{ab}$
    \item Non-degeneracy: For any $g,h\in\mathbb{G}$, the $e(g,h)\neq 1$ holds.
\end{enumerate}
\end{myDef}

\label{Sect:ShortSig}
Inspired by \cite{signature:boneh2004short}, we design the construction of the FedSOV short signature scheme as follows, and the four steps of the signature formalism are given in Sect. \ref{Sect:DigitalSignature}.
\begin{enumerate}
    \item $Setup(1^{\lambda})\rightarrow SP$. The $SP=(\mathbb G, \mathbb G_T, p, e, g)$. $\mathbb G$ and $\mathbb G_T$ are the bilinear elliptic group with prime order $p$. $e$ is the bilinear map, $g\in \mathbb G$ is the generator.
    \item $KeyGen(SP)\rightarrow(pk, sk)$. The $sk=(x,y)$ where $x$ and $y$ are randomly selected from $\mathbb Z^{*}_p$. The $pk=(u,v)$ where $u=g^{x},v=g^{y}$. The $sk$ is the credential, and the $pk$ is digested in the watermark.
    \item $Sign(m,sk)\rightarrow\sigma$. The signature $\sigma=(s,r)$, where $s=g^{1/(x+m+yr)}$ and $r$ is randomly selected from $\mathbb Z^{*}_p$.
    \item $Verify(m,\sigma,pk)\rightarrow(1/0)$. It verifies that 
    \begin{equation}
    \label{Equa:SigVer}
        e(\sigma, u\cdot g^{m}\cdot v^{r}) = e(g,g).
    \end{equation}
    If the Eq. \eqref{Equa:SigVer} holds, the algorithm outputs 1, then the ownership verification passes. Else the algorithm outputs 0, and the ownership verification fails.s
\end{enumerate}

\subsection{Experimental Setting Details}
\label{Sect:AppendixExperiment}
The detailed training parameters for federated model AlexNet, ResNet18 and DistilBERT are shown in Table \ref{Table:Traning Parameters}. The specific model architectures used in FedSOV are shown in Table \ref{Table:Alex} and Table \ref{Table:ResNet18}.

\subsection{Robustness against Targeted Destruction Attack}
\label{Sect:AppendixTarget}
Since the verification of FedSOV requires the public watermark and embedding matrix to support public verification, we further assume that the attacker has access to the watermark embedding position, embedding matrix, and watermark, and wants to destroy the integrity of the watermark by modifying the batch normalization parameters. This is a more targeted removal attack method than fine-tuning and pruning attack. Specifically, the attacker can add Gaussian noise approximating the parameter distribution to the parameters as follows:

For batch normalization parameters containing watermark information $\mathbf W_{\gamma}=(\gamma^{(1)},\cdots,\gamma^{(\omega)})$ with mean $\mu = \frac{1}{\omega} \sum\limits^{\omega} \limits _{i=1} \gamma^{(i)}$ and variance $\sigma ^2=\frac{1}{\omega} \sum\limits^{\omega} \limits _{i=1} {(\gamma^{(i)}-\mu)}^2$ , we add the Gaussian noise $\upsilon$ to $\mathbf W_\gamma$ as:

\begin{equation}
   \mathbf {W}_{err}=\mathbf W_\gamma+ \Upsilon    
\end{equation}
in which , ${\Upsilon}=(\upsilon ^{(1)},\cdots, \upsilon ^{(\omega)})$ , $\upsilon ^{(i)} \sim \mathcal N(\mu,\varphi \sigma^2)$, $\varphi  \in (0,1)$ is the noise intensity and $\mathcal N$ represents a Gaussian distribution. 

The trained ResNet18 with 1024 and 2048 bits watermark and DistilBert with 3076 bits watermark are used for this experiment. As shown in Fig. \ref{Fig:TargetedAttack}, it is impossible for an attacker to destroy watermarks in a targeted manner while maintaining model performance. When the attacker destroys the 1024 bits watermark of ResNet18/ 2048 bits watermark of ResNet18/ 3076 bits watermark of DistilBert in order to make the watermark detection rate lower than the security boundary, the performance of the model is not higher than 80/ 78/ 65 percent, resulting in at least 14/ 16/ 23 percent performance loss.

\subsection{Time Cost in Ownership Verification}
\label{Exp:TimeCost}
We implement the signature algorithm and test the performance of the algorithm by generating a 256-bit type-at elliptic curve group using the JPBC library under the equipment of Intel(R) Xeon(R) Gold 6133 CPU@2.50GHz. We use $mul$ for multiplication, $pow$ for exponential operations, $inv$ to denote the inverse operation and $pair$ for pairing operations, the time overhead of each type of operation under this curve is shown in Tab. \ref{Tab:BilinearTimeCost}.

\begin{table}[]
    \centering
    \caption{Time Cost of Bilinear Pair Operations}
    \begin{tabular}{|c|c|c|c|c|c|}
    \hline
        Operations & $\mathbb Z_p\ mul$ & $\mathbb Z_p\ inv$ & $\mathbb G\ mul$ & $\mathbb G\ pow$ & $\mathbb G_T\ pair$ \\
        \hline
        Time Cost & 12 $\mu$s & 20 $\mu$s & 48 $\mu$s & 11 ms & 18 ms\\
        \hline
    \end{tabular}
    \label{Tab:BilinearTimeCost}
\end{table}

We only need two addition operations, one multiplication operation, one inverse operation in $\mathbb Z_p$, and one exponentiation operation under the $\mathbb G$ to generate the signature, so the time overhead for signature generation is about 11 ms. Two exponential operations and two multiplication operations under $\mathbb G$ and one pairing operation under $\mathbb{G}_T$ are required to verify the signature, so the time cost for verifying the signature is about 40 ms. The time overhead of performing ownership verification is acceptable.

\end{document}